\documentclass[preprint2]{emulateapj}
\newcommand{\Lya}{Ly${\alpha}$}
\newcommand{\msun}{${\rm M}_{\odot}$}
\newcommand{\kms}{${\rm km\,s}^{-1}$}

\begin{document}
\title{The interstellar medium and feedback in the progenitors of the compact passive galaxies at z$\sim$2}

\author{Christina C. Williams\altaffilmark{1,2}, 
Mauro Giavalisco\altaffilmark{1}, 
Bomee Lee\altaffilmark{1},
Elena Tundo\altaffilmark{3},
Bahram Mobasher\altaffilmark{4},
Hooshang Nayyeri\altaffilmark{4},
Henry C. Ferguson\altaffilmark{5},
Anton Koekemoer\altaffilmark{5},
Jonathan R. Trump\altaffilmark{6,16}, 
Paolo Cassata\altaffilmark{7},
Avishai Dekel\altaffilmark{8},
Yicheng Guo\altaffilmark{9},
Kyoung-Soo Lee\altaffilmark{10},
Laura Pentericci\altaffilmark{11},
Eric F. Bell\altaffilmark{12},
Marco Castellano\altaffilmark{11}, 
Steven L. Finkelstein\altaffilmark{13},
Adriano Fontana\altaffilmark{11}, 
Andrea Grazian \altaffilmark{11},
Norman Grogin\altaffilmark{5},
Dale Kocevski\altaffilmark{14},
David C. Koo\altaffilmark{9},
Ray A. Lucas\altaffilmark{5},
Swara Ravindranath\altaffilmark{5},
Paola Santini\altaffilmark{11},
Eros Vanzella\altaffilmark{15},
Benjamin J. Weiner\altaffilmark{2}
}

\altaffiltext{1}{Department of Astronomy, University of Massachusetts, 710 North Pleasant Street, Amherst, MA 01003, USA, ccwilliams@email.arizona.edu}
\altaffiltext{2}{Steward Observatory, 933 N. Cherry Ave., University of Arizona, Tucson, AZ 85721, USA}
\altaffiltext{3}{INAF, Osservatorio Astrofisico di Firenze, Largo Enrico Fermi 5, I-50125, Firenze, Italy}
\altaffiltext{4}{Department of Physics and Astronomy, University of California, Riverside, 900 University Avenue, Riverside, CA 92521, USA}
\altaffiltext{5}{Space Telescope Science Institute, 3700 San Martin Boulevard, Baltimore, MD 21218, USA} 
\altaffiltext{6}{Department of Astronomy and Astrophysics, The Pennsylvania State University, University Park, PA 16802}
\altaffiltext{7}{Instituto de F\'isica y Astronom\'ia, Facultad de Ciencias, Universidad de Valpara\'iso, Gran Breta\~na 1111, Valpara\'iso, Chile} 
\altaffiltext{8}{Racah Institute of Physics, The Hebrew University, Jerusalem 91904, Israel}
\altaffiltext{9}{UCO/Lick Observatory, Department of Astronomy and Astrophysics, University of California, Santa Cruz, CA 95064, USA}
\altaffiltext{10}{Department of Physics, Purdue University, 525 Northwestern Avenue, West Lafayette, IN 47907, USA}
\altaffiltext{11}{INAF - Osservatorio Astronomico di Roma, via Frascati 33, 00040 Monte Porzio Catone, Italy}
\altaffiltext{12}{Department of Astronomy, University of Michigan, 500 Church Street, Ann Arbor, MI 48109, USA}
\altaffiltext{13}{Department of Astronomy, University of Texas, Austin, USA}
\altaffiltext{14}{Department of Physics and Astronomy, Colby College, Waterville, ME, 04901, USA}
\altaffiltext{15}{INAF - Osservatorio Astronomico di Bologna, Bologna, Italy}

\altaffiltext{16}{Hubble Fellow}

\begin{abstract}

 Quenched galaxies at z$>$2 are nearly all very compact relative to z$\sim$0, suggesting a physical connection between high stellar density and efficient, rapid cessation of star--formation. We present rest--frame UV spectra of Lyman--break galaxies (LBGs) at $z\sim3$ selected to be candidate progenitors of the quenched galaxies at $z\sim2$ based on their compact restframe-optical sizes and high $\Sigma_{SFR}$. 
We compare their UV properties to those of more extended LBGs of similar mass and star--formation rate (non-candidates). 
We find that candidate progenitors have faster bulk ISM gas velocities and higher equivalent widths of interstellar absorption lines, implying larger velocity spread among absorbing clouds. Candidates deviate from the relationship between equivalent widths of \Lya\ and interstellar absorption lines in that their \Lya\ emission remains strong despite high interstellar absorption, possibly indicating that the neutral HI fraction is patchy, such that \Lya\ photons can escape. We detect stronger \ion{C}{4}\ P-Cygni features (emission and absorption) and \ion{He}{2}\ emission in candidates, indicative of larger populations of metal-rich Wolf-Rayet stars compared to non-candidates. The faster bulk motions, broader spread of gas velocity, and \Lya\ properties of candidates are consistent with their ISM being subject to more energetic feedback than non--candidates. Together with their larger metallicity (implying more evolved star--formation activity) this leads us to propose, if speculatively, that they are likely to quench sooner than non--candidates, supporting the validity of selection criteria used to identify them as progenitors of $z\sim2$ passive galaxies. We propose that massive, compact galaxies undergo more rapid growth of their stellar mass content, perhaps because the gas accretion mechanisms are different, and quench sooner than normally--sized LBGs at these (early) epochs.

\end{abstract}

\section{Introduction}
\label{Introduction}

 The quenching of star-formation  is a major event 
in the evolution of passively evolving galaxies; on rapid timescales, the star--formation ends
 over the full volume of the galaxy
\citep{Renzini2006}, dust and cold gas contents are
removed, and the UV/optical SED evolves from blue
to red. 
The combination of these processes give rise to the well known Hubble
sequence, and galaxies appear to follow these characteristics even 10 Gyr ago
\citep{Franx2008, Kriek2009b, Toft2009, Wuyts2011, Cameron2011, Szomoru2011, Wang2012, BLee2013}. Yet, star-formation
quenching remains one of the most poorly understood physical processes
affecting galaxies at any epoch in the Universe's history. While there is
abundant evidence that both the baryonic and halo masses, and also the
environment of galaxies correlate with the cessation of star-formation
\citep{Kauffmann2003,Kauffmann2004, Baldry2004, Hogg2004, Blanton2005, Thomas2005, Peng2010,Peng2012,Woo2013,Tal2014}, there is a stark lack of
understanding of the physical processes that underpin these observed trends,
and how the fuel for star-formation, cold gas, in galaxies is affected.

Deep surveys such as the Great Observatories Origins Deep Survey \citep[GOODS;][]{Giavalisco2004} and the Cosmic Assembly Near-infrared Deep Extragalactic Legacy Survey \citep[CANDELS;][]{Grogin2011, Koekemoer2011} have, over the last decade, revealed a population of massive, 
 quenched
galaxies, very early in the universe's history  \citep[ z$\sim$2; e.g.][]{Franx2003, Cimatti2004, Glazebrook2004, Daddi2005, Trujillo2006, Bundy2006, Cimatti2006, vanderwel2008, VanDokkum2008, Fontana2009, Saracco2009, Bezanson2009, Cassata2011, Guo2012, Cassata2013}. 
Their spectra, colors and specific star-formation rates (sSFR) imply star-formation must have been quenched for at least $\sim$ 1 Gyr prior to observation \citep{Kriek2006, Kriek2009, Onodera2012, Kaviraj2012a, Gobat2012, vandeSande2013, Whitaker2013}. The remarkable thing about these quenched galaxies (besides the fact that they have already assembled such high stellar mass and also quenched star-formation only $\sim$2 Gyr after the Big Bang), is 
 that they are significantly more compact relative to local passive galaxies \citep[e.g.][]{Daddi2005,Zirm2007,Toft2007,VanDokkum2008, Damjanov2009,Buitrago2008,Muzzin2009, vanDokkum2010,Huang2013b, Davari2014}.
Nearly all ($> 80\%$) quenched galaxies at z$>$1.5 are more compact in stellar density than the lower 1$\sigma$ of passive galaxies at z$\sim$0 \citep[$\Sigma_{M*}>3\times10^{9}$ M$_{\odot}$kpc$^{-2}$;][]{Cassata2011,Cassata2013}. 
 More than 70$\%$ are ultra-compact ($\Sigma_{M*}>$1.2x10$^{10}$ M$_{\odot}$kpc$^{-2}$).  Very few local analogs to such galaxies exist \citep[][although see\citealp{Poggianti2013,Stockton2014,Trujillo2014}]{Trujillo2009,Taylor2010, ShihStockton2011}, and even intermediate redshift passive galaxies appear structurally different \citep{HsuStockton2014, Damjanov2014}. 
 Observations indicate this compactness is real, with large dynamical masses measured from velocity dispersions \citep{VanDokkum2009, Cappellari2009, Newman2010, vandeSande2011, Toft2012, Bezanson2013, Belli2014}, and deep high-resolution near-IR HST images and stacks confirming the small sizes 
 \citep{VanDokkum2008, Szomoru2012, Williams2014}.
 Such high masses and steep light profiles suggest that highly dissipative gaseous processes must have been behind the stellar mass assembly. 

 These galaxies pose extreme challenges to simulations of galaxy formation.
Simulated gas-rich mergers fail to reproduce the steep light profiles of these galaxies  \citep{Wuyts2010, Williams2014}, 
primarily because the stellar components in the merger progenitors cannot effectively dissipate angular momentum
 \citep{Hopkins2008a, Hopkins2013}. 
 Furthermore, observed starburst galaxies which are thought to be
gas-rich mergers due to their extreme star-formation rates (SFR), dust contents, and irregular
morphologies, appear morphologically distinct from the compact passive galaxies (CPGs), due to
their extended half-light radii and clumpy morphology \citep[][Giavalisco et
  al., in preparation]{Swinbank2010, Mosleh2011, Targett2011, Targett2012,
  Bussmann2012, Kartaltepe2012, Guo2012b}.
Transforming these extended starbursts into compact passive galaxies would require significant changes in morphology and gravitational potential within ~1 Gyr,
 and it is unclear what mechanism would be capable of
this. Alternatively, it has been argued that accretion of gas from the IGM
\citep[e.g.][]{BirnboimDekel2003, Keres2005, DekelBirnboim2006} can produce
compact galaxies through the dissipation of angular momentum, e.g. through
disk instabilities \citep{Dekel2009, Bournaud2011, Dekel2013}. These
gaseous disks may be large \citep[e.g. R$_{e}\sim$6-8 kpc][]{Dekel2009, Tacconi2010, Tacconi2013}, and therefore the star-formation timescale in the accreted
gas must be larger than the gas dissipation timescale in order to leave a
negligible stellar halo \citep{DekelBurkert2014}. 
Direct accretion of gas from the
IGM, however, could deposit onto galaxies
through filamentary configurations such that
angular momentum is largely canceled, allowing star formation to proceed in
very compact regions of galaxies \citep[][]{Sales2012, Johansson2012,
  Cen2014}. In this case massive compact galaxies would mark the sites of
galaxy formation by cold accretion at high redshift. The disappearance of
these objects, either star--forming \citep[e.g.][]{vanderWel2014, vanDokkum2014} or passive \citep{Cassata2011,Cassata2013,Damjanov2014} at low redshift is consistent
with the idea that cold accretion is shut down at $z\lesssim 2$
\citep{Keres2005,DekelBirnboim2006, Keres2009}.

The fact that these
massive and compact galaxies dominate the population of quenched
galaxies at high redshift \citep{Cassata2011,Cassata2013} strongly suggests a
physical connection between very high stellar density
and the effective and rapid 
cessation of star formation. Indeed, quiescence is observed to be highly correlated with compact morphologies \citep{Bell2012,Cheung2012}. This hypothesis is consistent with the observation that equally compact galaxies which still form stars virtually disappear from the universe by z$<$1, and has prompted analyses of compact star-forming galaxies (SFG) as progenitors of these CPGs based on a diversity of selection criteria \citep{Patel2013, Barro2013, Stefanon2013, Williams2014}. 
To what extent these studies overlap in their sample selections is unclear, and detailed analysis of the light profiles of the samples is necessary to understand if their stellar distributions are indeed consistent with that of the descendent CPGs.  The high fraction of AGN detection in some samples \citep[$\sim$50\%; e.g.][]{Barro2013} makes it unclear whether it is the AGN in the galaxy which is responsible for the quenching, or the compactness of the stellar distribution. \citet{DiamondStanic2012}, although studying compact SFGs at lower redshifts, find quenching is likely in compact SFGs without the necessity of AGN.

The implied evolutionary connection between compact SFGs and CPGs illustrates the need to study the properties of compact SFGs in detail. A recent study by \citet{Nelson2014} provided evidence for this evolutionary connection, with the discovery of a massive, compact SFG whose morphology, mass, and (gas) velocity dispersion match that of the CPGs, without indication of AGN activity. 
Recently, \citet{Williams2014} identified plausible ``candidate" progenitors of these CPGs among compact Lyman-Break Galaxies (LBGs) at z$\sim$3.4, based on their morphological similarity, and also their stellar masses and SFRs, such that they would be capable of matching the selection criteria of CPGs by the time the universe evolves to z$\sim$1.5. Their co-moving number density is similar to those of the CPGs, consistent with the suggestion that massive compact LBGs can evolve into the first quiescent systems, and the intensely star-forming, but much larger, sub-millimeter galaxies \citep[e.g.][]{Toft2014} are not the only avenue to such early quenching \citep{Stark2009, Williams2014}. These candidate progenitors are generally small in radius, and deep stacks show steep and compact stellar distributions, very similar to those of the CPG at z$\sim$2.
 Furthermore, their rest-frame UV SED appears redder compared to more normal SFGs (in terms of mass and size), while having consistent colors at rest frame optical-near IR \citep{Williams2014}. This shows that the redder UV indicated a more evolved stellar population than other LBGs.

These trends give rise to a number of unanswered questions. Could the
dense stellar distribution in CPGs have formed through some
dissipative gaseous process, 
 such as accretion of cold gas from
the intergalactic medium \citep[][]{BirnboimDekel2003, Keres2005,
  DekelBirnboim2006,Sales2012, Johansson2012,
  Cen2014}? What differentiates these compact galaxies such that
they are so small and dense relative to the rest of SFGs at early epochs? What
is the physical process that quenches their star-formation so early, and is it directly related to stellar density? 
Therefore, it is imperative to search for evidence (if any) that the physical conditions in these compact galaxies are particularly hostile to continued star-formation.

 The phenomenology of feedback in SFGs has been extensively documented \citep[e.g.][among others]{Shapley2003, Martin2005, Tremonti2007, Weiner2009, MartinBouche2009, Rubin2010, Steidel2010, Genzel2011, Heckman2011, SNewman2012, Rubin2014, ForsterSchreiber2014}. However, how feedback depends on various properties (e.g. SFR, stellar density, surface density of star--formation) remains to be fully empirically understood.
Feedback in the form of bulk outflows is observed ubiquitously in high-redshift SFGs, and high-velocity outflows have been observed in very compact starbursts, presumably due to the high surface densities of star-formation \citep{DiamondStanic2012}. 

Recently, \citet{Law2012b} studied the correlations between rest-frame optical morphology of galaxies, and their UV spectroscopic properties. The authors found that smaller galaxies (R$_{e}<$2 kpc) are more likely to exhibit both  \Lya\ in emission and more blue shifted interstellar absorption line centroids, suggesting a larger fraction of the gas appears outflowing in the smaller sized galaxies, as opposed to the larger fraction of gas at zero velocity in the larger galaxies. The presence of more zero velocity gas would imply that a larger fraction of the gas in large galaxies is not outflowing or as energized kinematically. 
 The interpretation is that the feedback in larger galaxies is less effective in energizing the ISM gas, thus increasing the optical depth of gas at rest with respect to the galaxy, bringing absorption line centroids closer to the systemic velocity and diminishing the ability for \Lya\ to escape \citep{Law2012b}. 

Motivated additionally by these recent findings that galaxy morphology correlates with spectroscopic properties such as velocity spread of interstellar absorption line centroids, with this study 
we aim to characterize the nature of the quenching mechanism(s) affecting
compact galaxies. This study presents a comparative analysis of spectroscopy
of  ``candidate" star-forming progenitors of CPGs, compared with the
properties of normal SFGs at the same epoch. 
 In Section \ref{Data}, we present the LBG samples
at $z\sim3.4$ and their spectroscopy, in Section \ref{Results} we present the
results of this comparative analysis, and in Section \ref{Discussion}, we
discuss the implications of our findings in the context of star-formation quenching at
high-redshift. Throughout this paper we assume a cosmology with
$\Omega_{\Lambda}=0.7$, $\Omega_{m}=0.3$, and $H_{o}=70$km s$^{-1}$
Mpc$^{-1}$.

\section{Data}
\label{Data}

\begin{figure*} [!!t]
\begin{center}
\includegraphics[scale=0.5]{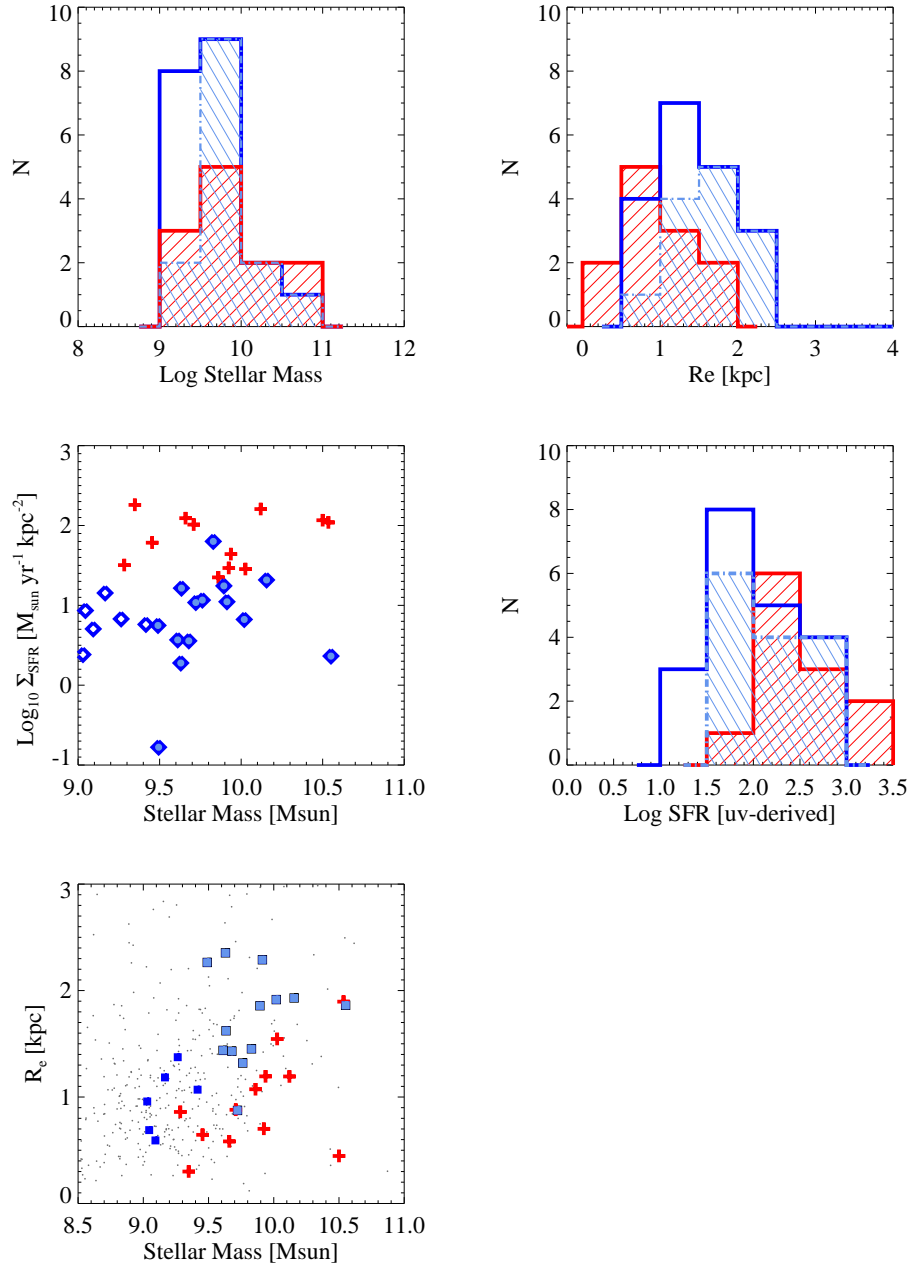}
\caption{Photometric and morphological properties of both good quality spectroscopic samples of candidates (red) and non-candidates (dark blue). In light blue (dot-dashed lines) are plotted the distributions of non-candidates once low-mass galaxies are removed, as discussed in the text. The mean masses of red and light blue samples are the same.  Histograms: The stellar mass distributions are very similar, while candidates exhibit higher SFRs and smaller half-light radii. 
As a result, the candidates have higher surface densities of star-formation (top right panel). Bottom right: candidates are more compact (i.e. smaller in size at a given stellar mass).
}
\label{sampleprop}
\end{center}
\end{figure*}

\subsection{The LBG samples}
The galaxies discussed here are the LBG samples at z$\sim$3.4 presented in \citet{Williams2014}. These LBGs are z-band selected from the the Great Observatories Origins Deep Survey \citep[GOODS;][]{Giavalisco2004}, and are found over an area of 113 arcmin$^{2}$ from the GOODS-South field {\it HST}/WFC3 imaging from CANDELS \citep{Grogin2011, Koekemoer2011}. 
 Through simulations we established our sample is 90$\%$ complete with z$\le$26.5 for galaxies of $<$0.3'' half-light radius \citep{KHHuang2013}. We measure multi-wavelength photometry using the object template-fitting method \citep[TFIT;][]{Laidler2007} software, which includes U--band imaging from the Visible Multiobject Spectrograph (VIMOS) on the Very Large Telescope \citep[VLT;][]{Nonino2009},  HST/ACS B,V,i,z-band, HST/WFC3
J, H-band, VLT/ISAAC K$_{s}$ photometry \citep{Retzlaff2010}, and {\it Spitzer}/IRAC 3.6, 4.5, and 5.7 $\mu$m imaging (M. Dickinson et al., in preparation). The details of the construction of multi-wavelength photometry are discussed in \citet{Guo2013}.
Measurements of galaxy properties for the LBG samples using the multi-wavelength photometry  assume a Salpeter initial mass function (IMF), and is described in \citet{Guo2012, Williams2014}.  Note that the specific assumption of an IMF here is
 relatively inconsequential, since the comparison between the two types of
 galaxies is done at fixed stellar mass. Changing the shape of the IMF will
 result in different absolute values of stellar mass, but comparisons
 between the two flavors of galaxies would remain unchanged. We use GALFIT \citep{Peng2002} to fit Sersic models to the WFC3 H-band images of the galaxies (rest--frame 3600 \AA  at z$\sim$3.4) from which we measure half-light radii (the sample are bright, H$<$25, which ensures robust morphological measurements). A detailed description of these measurements and also the selection of the LBG samples are presented in sections 2 and 3 of \citet{Williams2014}, and we briefly outline the procedure here.

This sample of LBGs was split into plausible ``candidate" progenitors of CPGs, and  ``non-candidate" LBGs as described in \citet{Williams2014}. The selection of candidate progenitors is based on evidence that any star-forming progenitors must also be compact, since merging and accretion tend to increase the sizes of galaxies \citep[e.g.][]{Hopkins2008a}. Therefore to identify candidate progenitors, we require that they be of similar size and morphology as the CPGs, and also have high enough SFR that if they quench soon after observation they will meet the CPG selection criteria \citep{Cassata2013}. This sample of CPGs is selected from CANDELS photometry to have low specific SFRs ($<$10$^{-2}$ Gyr$^{-1}$), high masses ($>$10$^{10}$M$_{\odot}$), non-detections in 24 and 100 $\mu$m maps inconsistent with dust obscured SF, and visually identified in the H-band to be ''early-type". 
 It is impossible to know the future star-formation histories of galaxies; however, CPGs have strong constraints on their star-formation histories (SFH), because their low sSFRs implies they must have quenched star-formation $>$1Gyr prior to the epoch of observation  \citep{Kriek2006, Kriek2009, Onodera2012, Kaviraj2012a, Gobat2012, vandeSande2013, Whitaker2013}. Therefore, we assumed a projected (future) star-formation history for all galaxies \citep[exponentially declining with $\tau$=100Myr so that all galaxies would be passive according to the sSFR criteria of][]{Cassata2011,Cassata2013} with the observed initial SFR, and estimated how much stellar mass would be accumulated from the time of observation until the epoch of CPGs, which we took to be $<z>$=1.6. We additionally assumed no growth in size. Those LBGs which would meet the CPG selection of \citet{Cassata2011,Cassata2013} are considered candidate progenitors (from now on referred to as candidates). The candidate selection is primarily based on SFR and size, and to a lesser extent, the stellar mass. In other words, candidates tend to be more compact and a milder tendency to higher SFR.  The combination of these two characteristics means that candidates have higher surface density of SFR ($\Sigma_{SFR}$; see Figure \ref{sampleprop} for the properties of the subset of candidates and non-candidates with spectroscopy studied in this paper). The average SFRs, stellar masses, and circularized half-light radii, for candidates (non-candidates) in the spectroscopic sample are 520 (170) M$_{\odot}$yr$^{-1}$,  9.8 (9.6) log$_{10}$M$_{\odot}$, and 0.9 (1.9) kpc.

\begin{figure*} [!!t]
\begin{center}
\includegraphics[scale=0.8, trim=6cm 3cm 6cm 3cm]{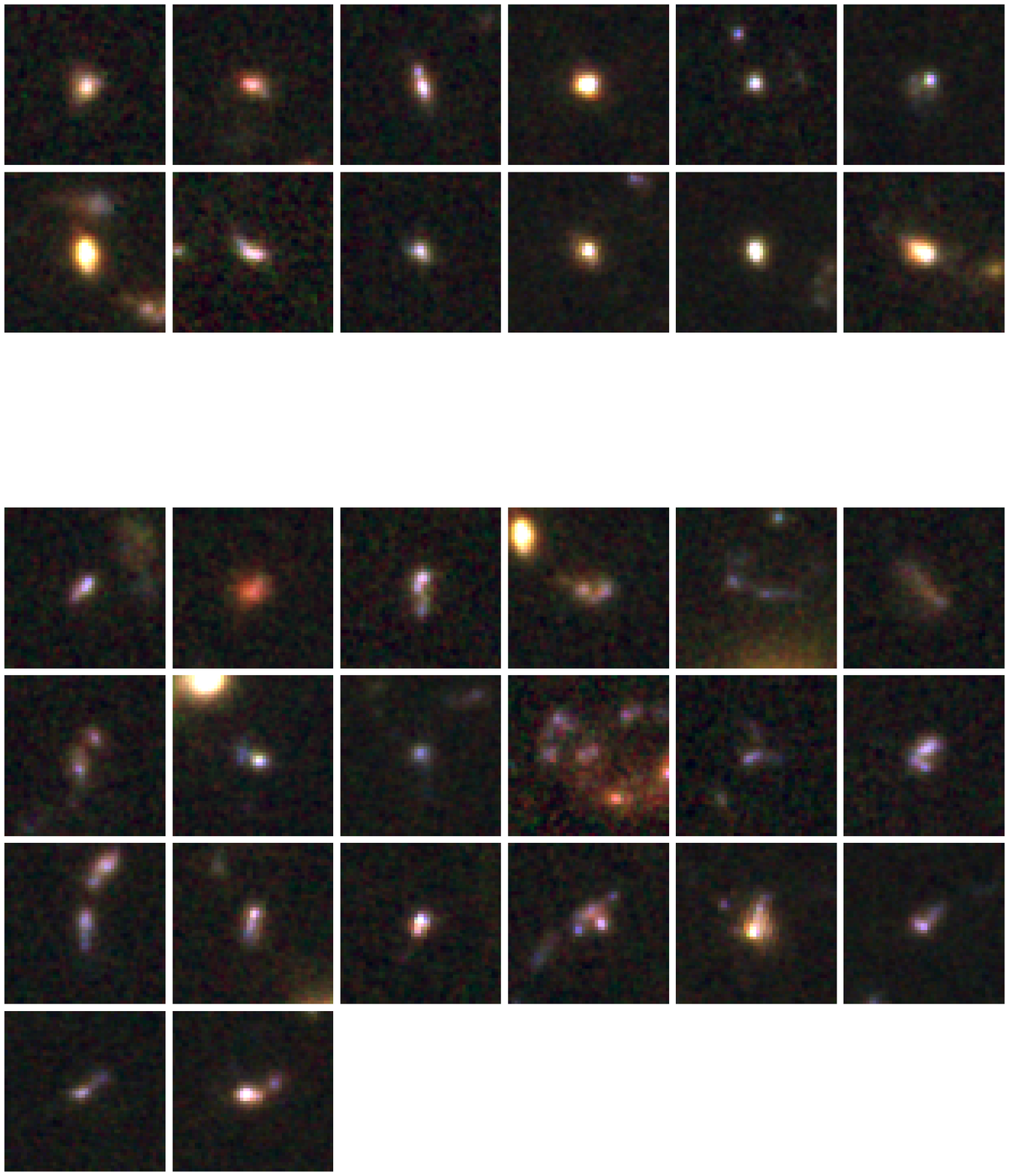}
\caption{Color composites from the ACS and WFC3 filters, of candidates (top) and non-candidates (bottom). Images are 2 arc seconds on a side.}
\label{cutouts}
\end{center}
\end{figure*}

\subsection{Spectroscopy}
We use spectra in GOODS-South obtained at the European Southern Observatory's Very Large Telescope (ESO/VLT) as part of the GOODS spectroscopy program with FORS2 \citep{Vanzella2005, Vanzella2006, Vanzella2008}, VIMOS \citep{Popesso2009}, and the Galaxy Mass Assembly Ultra-deep Spectroscopic Survey with FORS2 \citep[GMASS;][]{Kurk2009,Kurk2013}.  At z$\sim$3.4, these optical spectra probe the rest frame UV spectrum of the LBGs, a spectral region rich with diagnostics of the physical state of the ISM. The spectral resolution of these surveys varies between R$\sim$250-1000, corresponding to velocity resolutions of 300-1200 \kms. The majority (60$\%$) of the spectra we use in our analysis have R$\sim$660-1000 (velocity resolution 300-450 \kms).

\begin{figure*} [!!t]
\begin{center}
\includegraphics[scale=0.32]{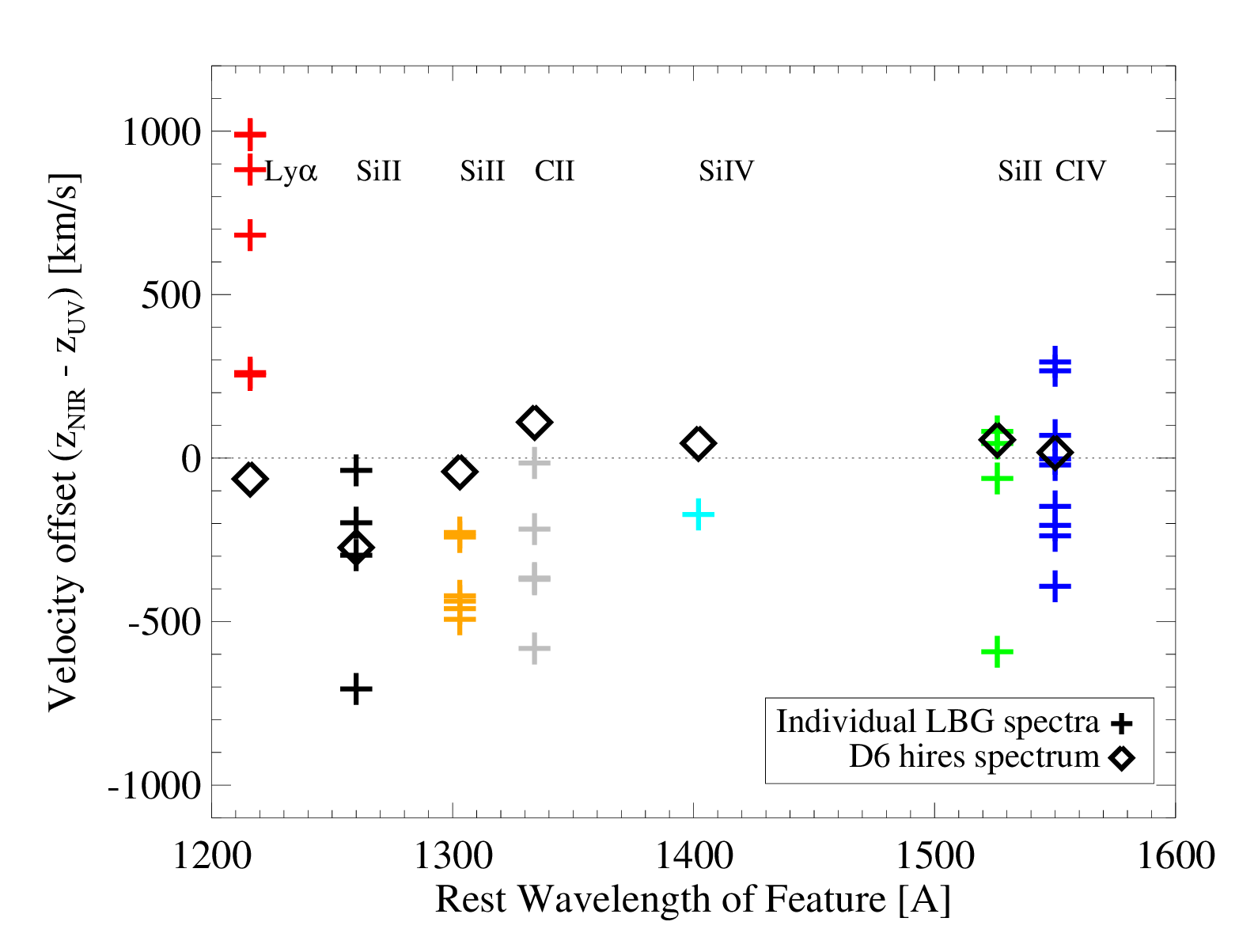}
\includegraphics[scale=0.32]{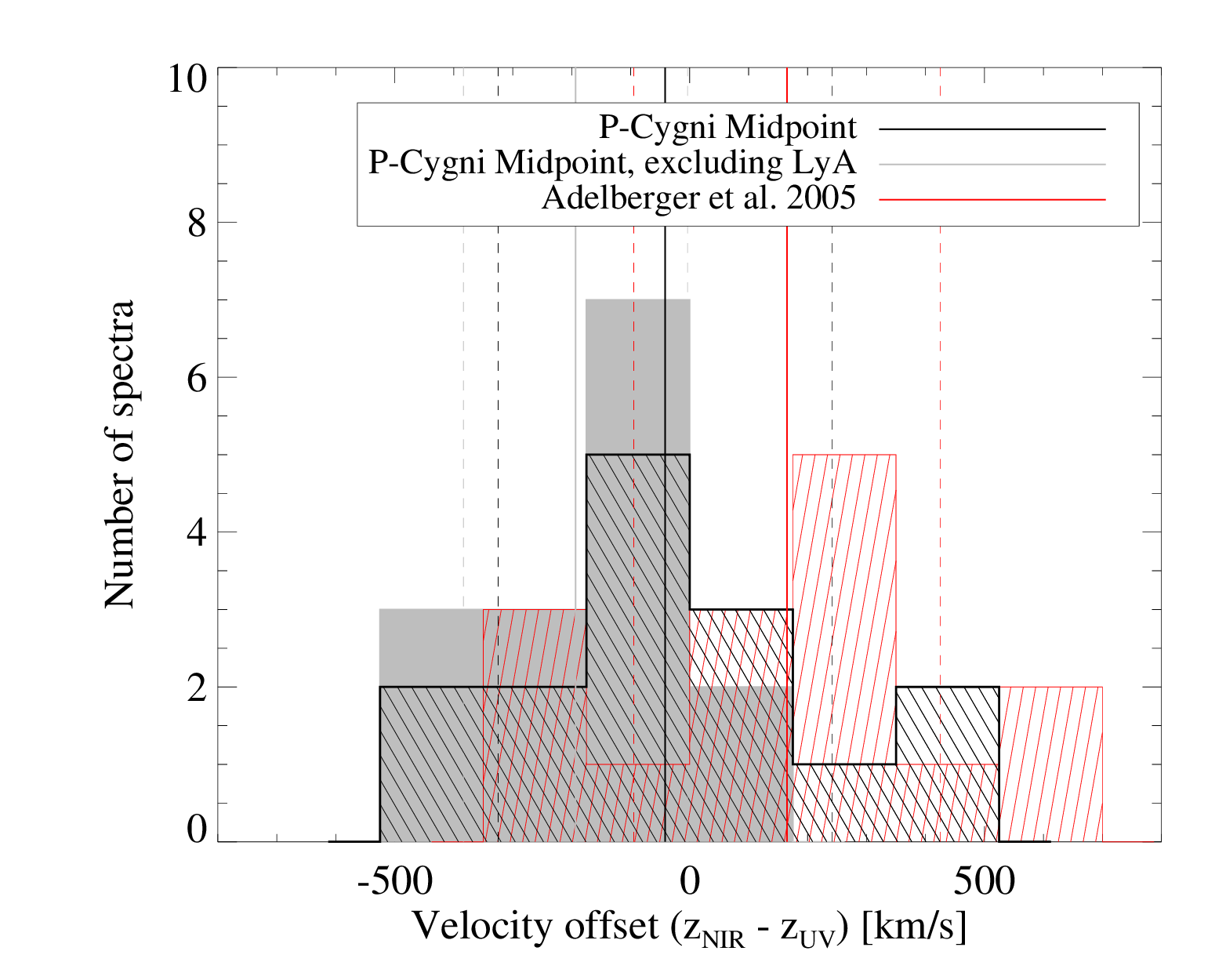}
\caption{Left Panel: Velocity offset between the midpoint method of deriving the systemic redshift (as described in the text) and the actual systemic velocity from NIR spectroscopy, for individual galaxy spectra. Right Panel: Distribution of velocity offsets when individual midpoint-derived redshifts are averaged (black), which has a mean velocity offset of -42 (solid line) and a standard deviation of 283 \kms (dashed lines). In grey are the same measurements when Ly$\alpha$ is excluded from the measurement, whose distribution has a larger mean offset of -193, and standard deviation of 190 \kms. In red is the same when measured using the method of \citet{Adelberger2005}, which has a mean offset of 165 \kms with standard deviation of 260 \kms. }
\label{systemic}
\end{center}
\end{figure*}

The scientific goal of the surveys above were to identify the redshifts of large samples (tens of thousands) of galaxies at z $>$1, and while there are often multiple identifiable emission or absorption lines, generally the spectra do not have adequate signal-to-noise ratio for detailed studies of the galaxy properties on an individual basis. In the following sections we therefore study the average properties using stacked spectra to maximize signal to noise.
 We visually inspected each individual spectrum, retaining only those of high quality for our final spectroscopic sample. This quality screening is qualitative, including visual identification of absorption and emission lines at the catalogued redshift, and also to some extent taking into account the quality flag published by each survey, whose redshifts may have been estimated using cross-correlations with galaxy templates rather than obvious lines. We additionally require ``good" spectra to have identifiable features which we use to estimate systemic redshifts (see next paragraph).
 Our final ''good'' quality spectroscopy sample includes 12 candidate progenitors and 20 non-candidates (27$\%$ and 15$\%$ of the original samples from \citet{Williams2014}, respectively).   The total exposure time in the stacks of each are $\sim$103 and $\sim$98 hours, respectively.

\subsection{Systemic Redshifts}
\label{syst}
The catalogued redshifts that are published in the spectroscopic surveys were typically measured using spectral lines which are either resonant lines (Ly$\alpha$) or strong interstellar lines which may be heavily affected by outflows (for example \ion{Si}{2}, \ion{Si}{4} and \ion{C}{4}). Therefore these redshifts are unlikely to be systemic (tracing the exact redshift of the galaxy) 
but rather reflect the redshift of the galaxy, modulated by additional velocities due to the kinematic properties of the ISM.
 Interstellar absorption lines are often observed blue shifted with respect to the rest transition wavelength due to large-scale outflows in the ISM of high-redshift galaxies. Ly$\alpha$ emission is often observed redshifted, since when redshifted out of resonance Ly$\alpha$ photons may travel freely, and therefore the \Lya\ may also reflect the properties of outflows \citep{Shapley2003, HansenOh2006, Verhamme2008, SchaererVerhamme2008, Vanzella2009, Steidel2010, Kulas2012}.  Lines which trace the systemic redshift, for example photospheric lines which originate in the atmospheres of stars, are generally too faint to be detected in our spectra.

 In this section, we document our procedure to estimate systemic redshifts using the UV spectroscopy. We also quantify our expected uncertainty in the systemic redshifts using a subset of galaxies for which systemic redshifts have been published based on rest-frame optical nebular emission lines, using NIR spectra. These NIR spectra and systemic redshifts come from Keck/MOSFIRE \citep{Holden2014} and the Assessing the Mass-Abundence redshift[-Z] Evolution (AMAZE) survey \citep{Maiolino2008} with VLT/SINFONI. We use a comparison of 15 UV spectra of LBGs with published NIR-derived redshifts.

Previous studies have published methods to derive systemic redshifts from rest-frame UV spectroscopy, when no systemic lines are available \citep[e.g.][]{Adelberger2005, Steidel2010}. In particular, \citet{Adelberger2005} have used a large sample of LBG at z$\sim$3 with near-IR spectra to estimate systemic redshifts from the UV lines only, for the case where only \Lya\ is present, both ISM lines and \Lya\ are detected, and when only ISM lines are detected. This method is based primarily on the observed offsets in velocity between the various UV lines and NIR observed nebular emission lines. These velocity offsets are direct measurements of velocities in the ISM, and therefore these conversions from UV-line redshifts to estimated systemic redshifts will reflect the average outflowing ISM properties of their LBG samples. This means, by using their conversion, we artificially imprint the average kinematic properties of their LBG samples on our two galaxy samples, and as we are interested in studying outflow velocities, we approach this method with caution. We seek the best method to derive systemic redshifts from our samples, without any assumptions about the properties of the outflowing ISM.

Our individual spectra often exhibit emission lines immediately redward of broad interstellar absorption. 
 These are most often seen in \ion{C}{4}$\lambda\lambda$1548,1550 and Ly$\alpha$, which often exhibit P-Cygni profiles, but are also observed in other absorption lines (e.g. \ion{C}{2}$\lambda$1334, \ion{Si}{2}$\lambda$1526). 
Motivated by the need for an entirely empirical systemic redshift estimator that is independent of assumptions about the average properties of LBGs, we test how the redshift measured by the transition point between the absorption and emission components in various elemental species relates to the NIR-derived systemic redshift where available.

We measure these redshifts using a method based on the midpoint between the centroid wavelengths of emission and absorption components. We measure the wavelength of the midpoint between centroids for each observed feature, assuming this ``midpoint" wavelength to be at rest with respect to the galaxy, and measure its redshift. Note that we are not suggesting a physical scenario to motivate the use of this mid-point method to estimate systemic redshifts and rather the method is entirely empirical. The left panel of Figure \ref{systemic} shows the velocity offset between the predicted systemic redshift from the most common individual features (e.g. Ly$\alpha$, \ion{Si}{2}$\lambda$1260, \ion{O}{1} / \ion{Si}{2}$\lambda$1303, \ion{C}{2}$\lambda$1334, \ion{Si}{2}$\lambda$1526, \ion{C}{4}\ $\lambda\lambda$1548,1550) compared with the systemic redshift of the galaxy measured from the NIR. We supplement these measurements with measurements from the Keck/HIRES spectrum of D6 \citep[N. Reddy, C. Steidel, personal communication;][]{Shapley2003,Steidel2010}, whose systemic redshift has been derived from NIR spectroscopy \citep{Pettini1998}. We find that Ly$\alpha$ is the poorest estimator of systemic redshift among our LBG sample, not surprising since the resonant nature of the escape of Ly$\alpha$ photons from a galaxy may mean its transition point is not at rest, and also due to the complication of identifying the absorption component due to the Ly$\alpha$ forest. The measure from the Ly$\alpha$ in the HIRES spectrum of D6 may be better due to the more accurate identification of the P-Cygni absorption component. We also find that transitions of \ion{Si}{2} may be poor estimators, possibly due to confusion between P-Cygni emission (if it exists) with fluorescent emission \ion{Si}{2}* (if present), which exists at slightly longer wavelengths than would be expected of P-Cygni emission. \ion{C}{4}\  appears to be a good estimator. The midpoint derived redshift from \ion{C}{4}\  is relatively accurate, with the mean systemic redshift centered at zero and never deviating more than 500 \kms from the rest transition, and is the most frequently observed in our spectra of all features tested here.

In the right panel of Figure \ref{systemic}, we show how the combined information of all available transitions in individual spectra (i.e. the average redshift from each available midpoint) compares with the NIR-derived systemic redshift. The average redshift appears to agree remarkably well with the NIR-derived one. The mean of the distribution of velocity offsets between midpoint-derived redshifts and systemic redshifts is -42 \kms, with standard deviation of 283 \kms. No galaxy deviates larger than 490\kms. Since Ly$\alpha$ is the most deviant in the left panel, we repeat the distribution of velocity offsets with the Ly$\alpha$ midpoint redshift removed from the average. We find that although removing Ly$\alpha$ tightens the distribution to 190 \kms, the offset from 0 increases to a mean offset of -193 \kms. 
We therefore conclude that we do not gain a closer estimate to systemic by excluding the Ly$\alpha$, and it should be included where available.

\begin{figure*} [!!t]
\begin{center}
\includegraphics[scale=0.45]{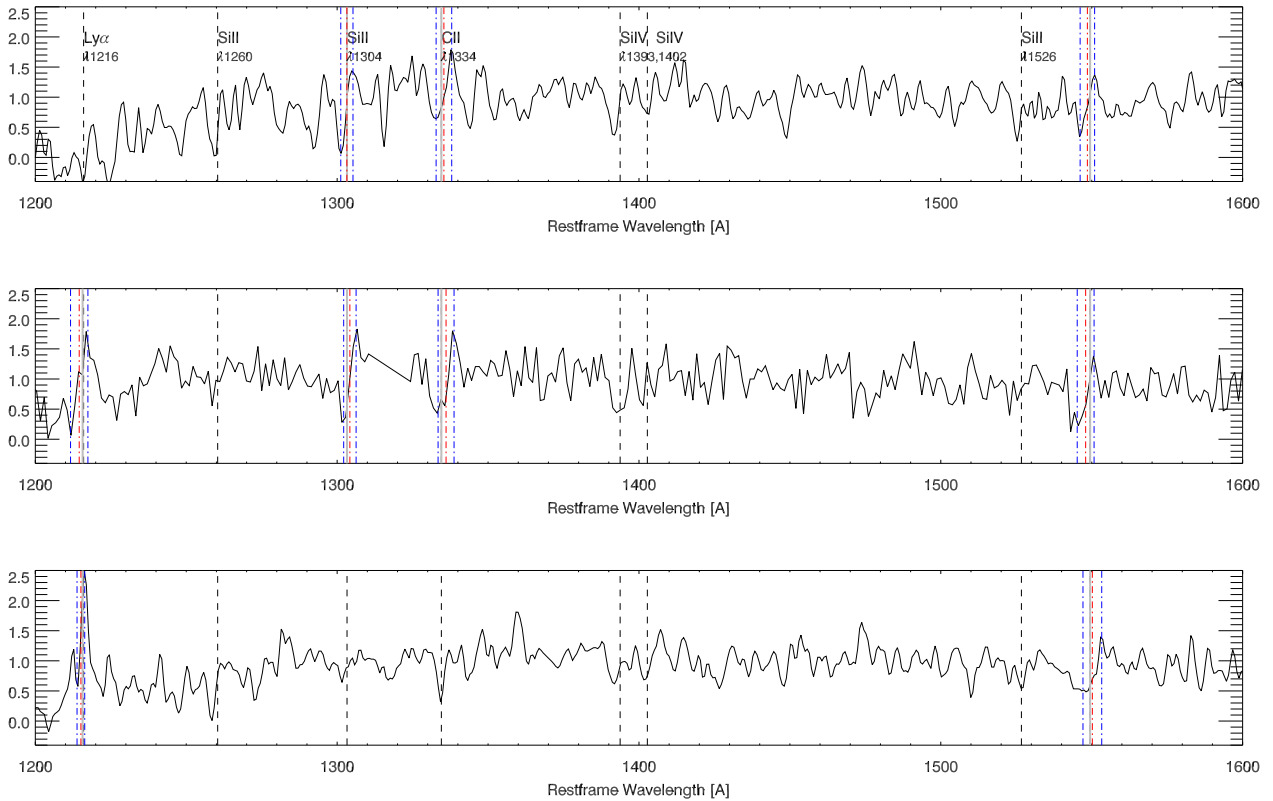}
\caption{ Individual VLT spectra that go into the stack. For each spectrum, the lines used in the calculation of the systemic redshift are identified, with blue showing the peak and trough and red showing the calculated midpoint. Grey vertical lines show the rest-wavelength of the line used for estimating the systemic redshift after de-redshifting the spectrum with the average of the red lines. The rest frame wavelength of other interstellar lines that are not used to estimate redshift are shown with black dashed lines. }
\label{refspec}
\end{center}
\end{figure*}

In addition to testing the midpoint between emission and absorption, we also tested using the transition point between where the emission and absorption components meet the continuum level, and found that this continuum transition point
did not trace systemic as accurately as the midpoint, and this may be due to the difficulty in accurately measuring the continuum level in noisy spectra.

For comparison, we have additionally computed systemic redshifts estimated using the \citet{Adelberger2005} method for comparison with the observed NIR redshifts. These are also plotted in red in the right panel of Figure \ref{systemic}, and find that the \citet{Adelberger2005} have a mean offset of 165 km s$^{-1}$, with a standard deviation of 260 km s$^{-1}$. This average offset is larger than that derived using the mid-point method with a similar distribution width, and results in multiple galaxies with offsets larger than 550 km s$^{-1}$. 

Therefore, in the remainder of our analysis, we estimate systemic redshifts based on the midpoint method we have outlined here, which we have characterized to be a tracer of systemic with an expected uncertainty of $\sim$280 \kms. Since the main purpose of this paper is to study the {\it relative} properties of candidate progenitors, compared to ordinary LBGs, it is therefore very important to use a uniform way to estimate redshift from all galaxies studied here, {\it independent} of any assumptions on the properties of outflows in the ISM \citep[e.g.][]{Adelberger2005,Steidel2010}. Although individual outflow velocities may not be accurate to more than this velocity uncertainty, we choose this option rather than mixing redshift estimators for different galaxies, which could smear away differences between the samples. Importantly, however, we found that the results of this paper remain if we instead adopt either the \citet{Adelberger2005} method, or, if we combine the mid-point method with the NIR-derived redshift where available when producing the stacks. We found that {\it all qualitative comparisons of the stacks are independent of the choice of the above methods} used to convert spectra to the rest frame, and we would reach the same conclusions were we to adopt any of them.

\subsection{Stacking Procedure}
\label{stackproc}
We stack our two samples of galaxies using the following procedure. We first individually de-redshift each spectrum using the estimate of systemic redshift from the mid-point of P-Cygni-like profiles discussed above. We then normalize each spectrum using the median value of the continuum measured between rest frame 1400 $< \lambda <$ 1500{\AA}. Finally, we stack using the {\em scombine} package in IRAF, where we rebinned to a common dispersion 
 ($d\lambda$ = 0.6{\AA} pixel$^{-1}$, rest frame), and performed a 3$\sigma$ clipping during the stack (although we note that the clipping minimally affected the resulting stack, indicating that there are few significant outliers in terms of the features within our samples). Our final stacks are presented in the top panel of Figure \ref{spec}. The bottom panel shows the spectral ratio (candidate stack divided by non-candidate stack). The ratio should be unity for identical features between the two samples, and therefore large deviations from this indicate features which differ. An indication of significance of the spectral differences can be see in the pink region, which is defined by unity plus or minus the combined (in quadrature) sample standard deviations of candidates and non-candidates. These standard deviations are estimated as follows. We repeat the stacks of each sample using jackknife resampling, each time removing one spectrum and stacking the rest. The final sample error on each spectral point is then the standard deviation of the jackknifed stacks. We then combine in quadrature the standard deviations of candidates and non-candidates around unity to make the pink region. We note that the combined region redward of 1250{\AA} is remarkably uniform, indicating the variation in those spectral features from galaxy to galaxy is relatively small. In contrast, the pink region near Ly$\alpha$ is much larger, indicating a diversity of Ly$\alpha$ properties among our samples. The differences in stacked Ly$\alpha$ between the candidates and non-candidates, along with error bars indicating the standard deviation of each stack, can be seen in Figure \ref{lya}.

\section{Results}
\label{Results}
 \subsection{Sample Properties}
 \label{sect3.1}
\begin{figure*} [!!t]
\begin{center}
\includegraphics[trim=2cm 0cm 0cm 0cm,width=7.1in] {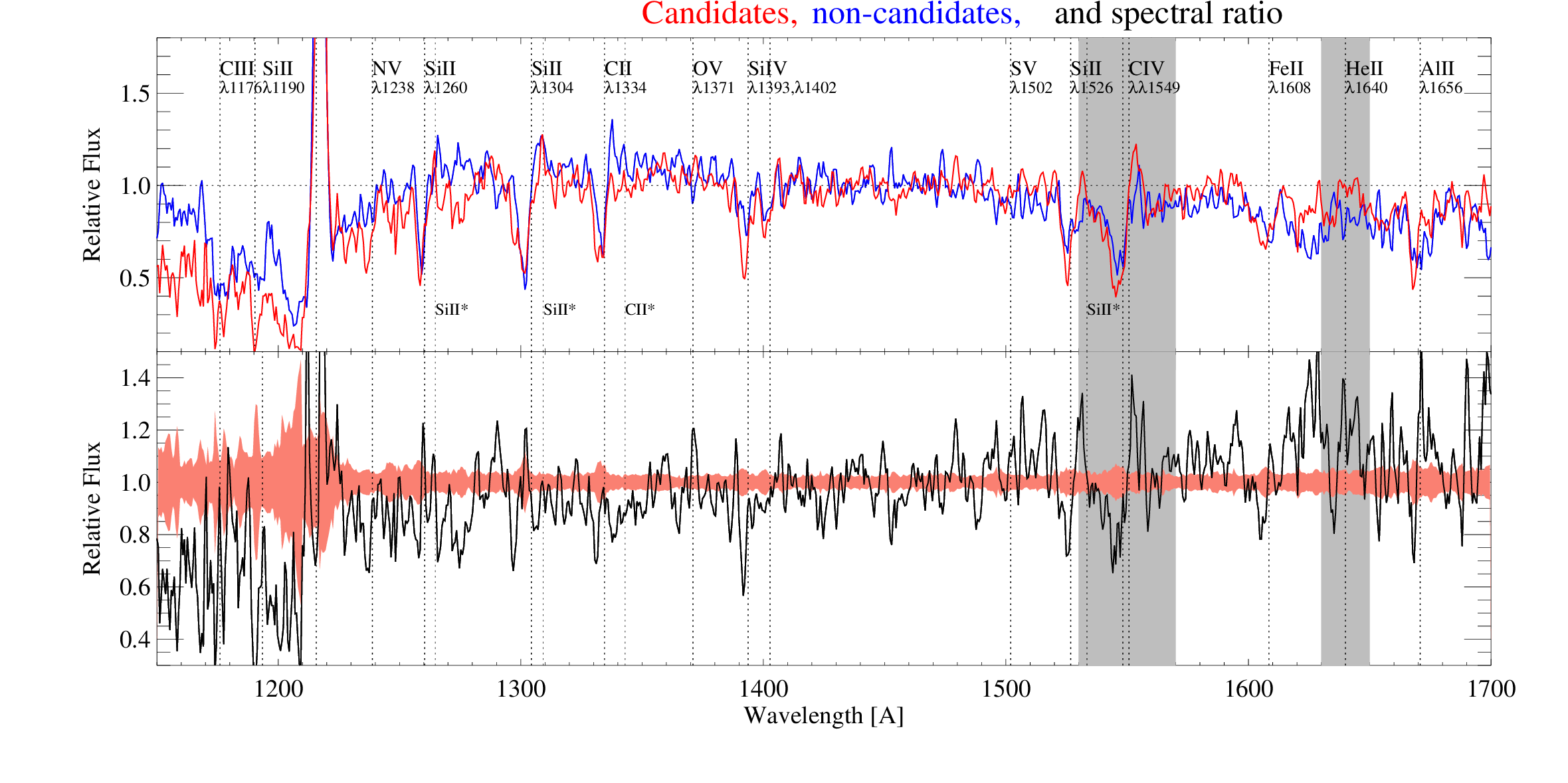}
\caption{Top panel: Stacks of good quality spectra for candidates (red), all non-candidates (dark blue). The rest frame wavelengths of various observed interstellar lines are identified with the dashed lines. Bottom panel: spectral ratio of the two stacks (i.e. candidate stack divided by non-candidate stack). Various features which differ between the two samples are highlighted by the ratio, which would be unity for identical features. Pink region indicates unity plus or minus the sample standard deviations of the two samples added in quadrature. Shaded regions indicate two features which depend on metallicity in rest frame UV spectra that will be discussed in section \ref{relmet}.}
\label{spec}
\end{center}
\end{figure*}

\begin{figure} [!t]
\begin{center}
\includegraphics[trim=2cm 0cm 0cm 0cm,scale=0.45]{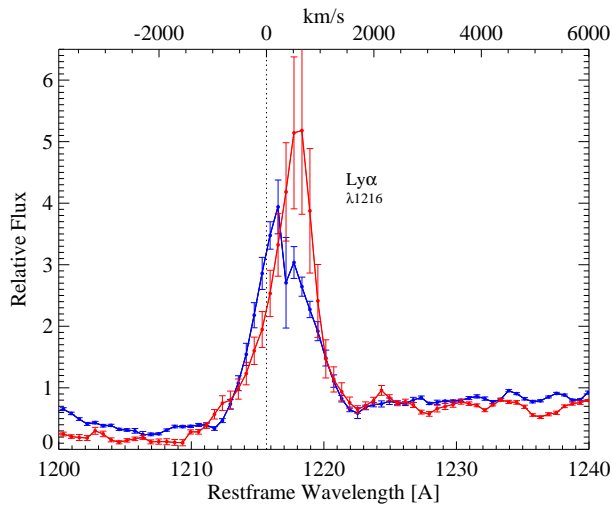}
\caption{Zoom in of the stacks (candidates in red and non-candidates in blue) presented in Figure \ref{spec} around \Lya\ .}
\label{lya}
\end{center}
\end{figure}

In the composite spectrum shown in Figure \ref{spec} there are many features that are generically visible among LBGs at z$\sim$3-4 \citep[e.g.][]{Shapley2003,Jones2012}. There are a variety of high and low ionization lines visible in absorption. First, by comparing the rest-wavelength of these lines (dotted lines) it is obvious that both candidates and non-candidates exhibit outflows, as indicated by the blue shifts in absorption troughs relative to the rest wavelength. In particular, \ion{Si}{2}, \ion{C}{2}, \ion{Si}{4},\ion{C}{4}\ , and \ion{Fe}{2} are visible in both spectra. We also see \ion{He}{2}\  in emission, detected in the candidates. The UV spectral slope 
 is redder for candidates than non-candidates, similar to the redder UV observed photometrically in \citet{Williams2014}. Both of the samples show strong \Lya\  in emission (although it extends beyond the range shown in Figure \ref{spec} to emphasize the weaker features in other parts of the UV spectrum, we show \Lya\ individually in Figure \ref{lya}). The candidates exhibit stronger absorption in \Lya\ , and their \Lya\ emission is also more redshifted. 
  As these are stacks and therefore represents the average properties of galaxies selected according to our criteria, we do not focus on specific quantities from each stack. Rather, in the following sections, we will focus on comparing the spectral properties in a {\it relative} way.

\begin{figure*} [!!t]
\begin{center}
\includegraphics[trim=2cm 0cm 0cm 0cm,width=7.1in] {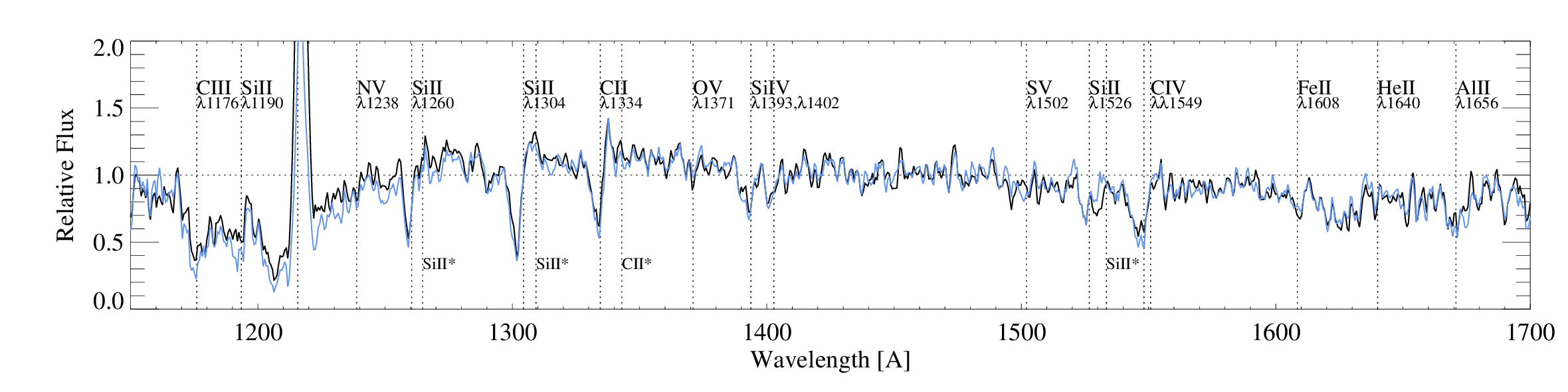}
\caption{Stack of all non-candidates as in shown in Figure \ref{spec} (black), and the stack of the high-mass sample of non-candidates (blue). Interstellar absorption line properties of the high-mass stack are very similar to that of the full sample. The major differences are a decrease in Ly$\alpha$ emission and redder far-UV slope in the high-mass sample.}
\label{highmass}
\end{center}
\end{figure*}

Because of this relative comparison, it is necessary to test if mass differences between the two samples could produce the relative differences we observe. For example, it is  
possible that the kinematic properties of the absorbing gas may be a function of luminosity or mass. 
 Additionally, if the mass-metallicity relation persists to z$>$3 \citep[e.g.][]{Maiolino2008}, samples which are offset from each other in mass may be more likely to show offsets in metallicity. Both \citet{Steidel2010} and \citet{KSLee2013} have found that velocity offsets between interstellar absorption lines and the systemic redshifts of galaxies are dependent on the baryonic mass of the galaxy. 
In the top left panel of Figure \ref{sampleprop}, the non-candidate sample has a larger low-mass tail than the candidates, although their overall mass range and high-mass distribution are very similar. To explore the dependence of any of our main conclusions on this low mass tail, 
we repeat the non-candidate stack with a subsample of non-candidates which are more similar to candidates in mass, such that the average mass of the distributions match.
 We remove all non-candidates whose log$_{10}$ M$^{*}$ $<$ 9.6, resulting in a sample of non-candidates with the same average stellar mass as the candidate sample, $\langle M_{\rm{star}}/M_\odot \rangle \sim 9.8$. This high-mass sample includes 14 of the 20 non-candidates. Due to the smaller number of non-candidates which meet this criteria in mass, the noise in the stack is larger.  The second (lighter) blue distributions in each panel of Figure \ref{sampleprop} shows the sample differences when excluding these lowest mass galaxies. The stack of the high-mass sample compared with the full sample of non-candidates is presented in Figure \ref{highmass}. We find a slight increase in the equivalent width of some of the interstellar absorption lines (e.g. \ion{Si}{4}$\lambda$1393), and a decrease in Ly$\alpha$ emission strength,  but otherwise, the high-mass stack is qualitatively very similar to that of the stack of the full sample of non-candidates. We stress that the main results of this study, reported in the next sections, remain unchanged if we used measurements made with the high-mass stack of the non-candidate galaxies. 
Therefore, our results are independent of any offsets in their mass distributions, and rather reflect intrinsic differences between the candidates and non-candidates.

\begin{figure*} [!t]
\begin{center}
\includegraphics[trim=2cm 0cm 0cm 0cm,width=7.1in]{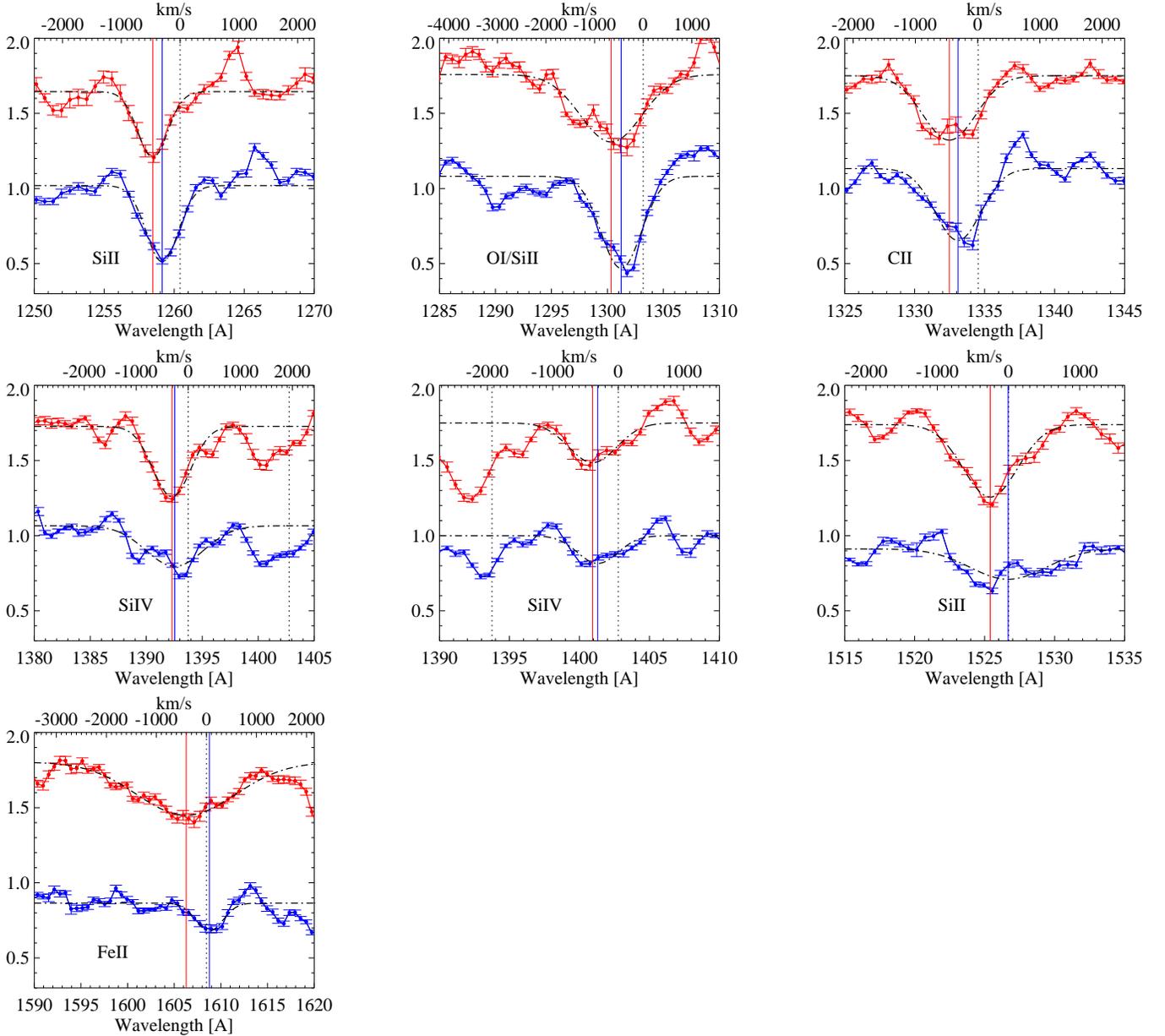}
\caption{Zoom-ins of various interstellar lines visible in candidate (red) and non-candidate (blue) spectra in Figure \ref{spec} (candidate spectra are arbitrarily offset by 0.8 for clarity). Gaussian fits to each line are shown as black dot-dashed lines. Fitted line centroids are shown as vertical lines of appropriate color. }
\label{spec_zoom}
\end{center}
\end{figure*}

\begin{figure*} [!t]
\begin{center}

\includegraphics[trim=2cm 0cm 0cm 0cm,width=7.1in]{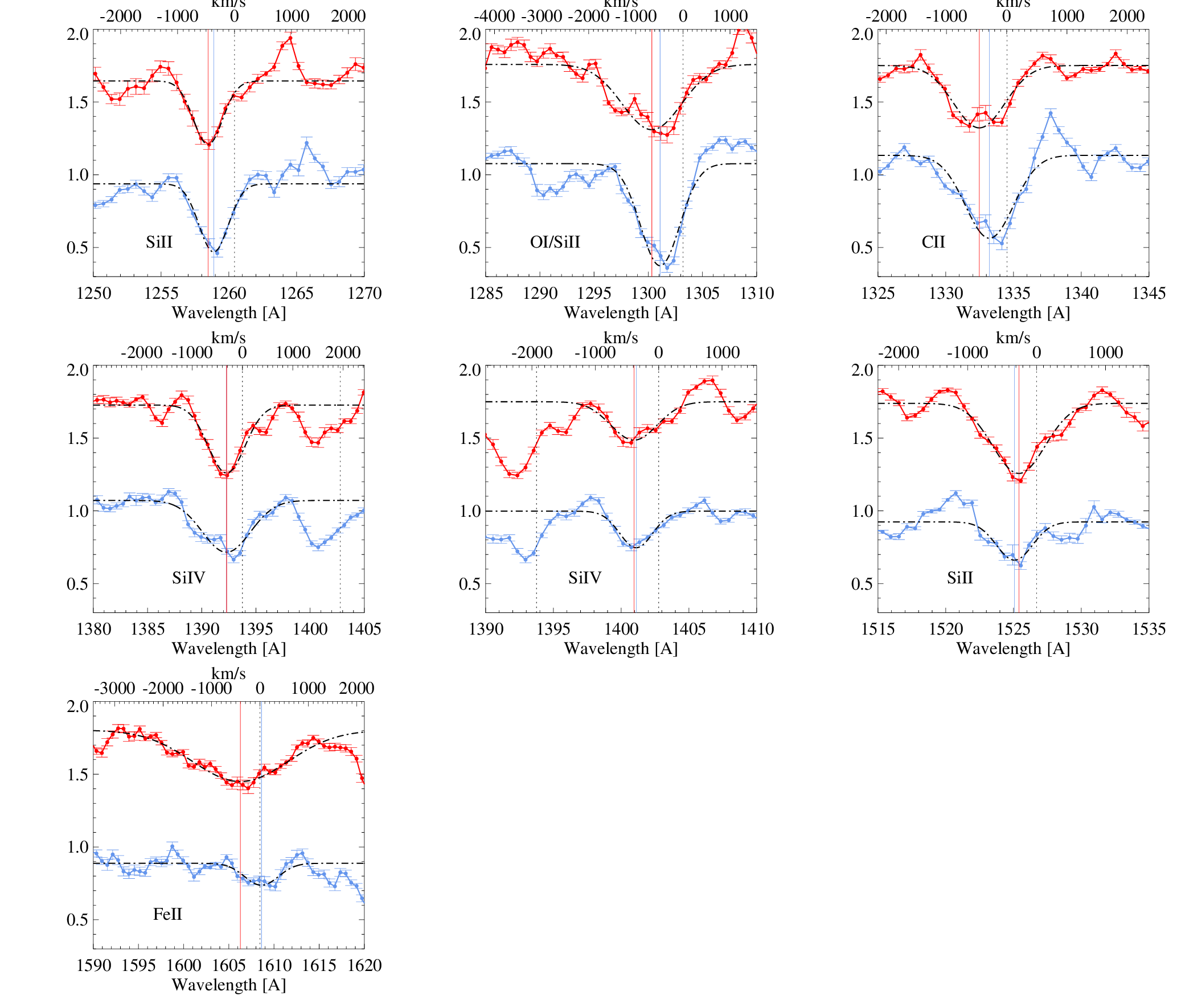}
\caption{Same as Figure \ref{spec_zoom}, but instead the blue spectra are from the high-mass non-candidate stack. The  centroid velocities in candidates are larger than both the full non-candidate sample in Figure \ref{spec_zoom} and high-mass non-candidate samples.}
\label{spec_zoom_hm}
\end{center}
\end{figure*}

 \subsection{Kinematic properties}
 \label{kinprop}
In this section, we focus on the relative kinematic properties of the ISM in the two samples of galaxies, which can be derived from the following observed interstellar absorption lines: \ion{Si}{2} $\lambda$1260, the \ion{O}{1}/ \ion{Si}{2} $\lambda$1303 blend, \ion{C}{2} $\lambda$1334, \ion{Si}{4} $\lambda$1393 and $\lambda$1402, \ion{Si}{2} $\lambda$1526, and \ion{Fe}{2}$\lambda$1608. As can be seen in Figure \ref{spec}, the candidates exhibit stronger absorption in the majority of these lines. Visual inspection of several lines shows the candidates have larger velocity offsets. Both of these are also true of the \ion{C}{4}\  $\lambda\lambda$1548,1550 doublet, however due to the asymmetric nature of the P-Cygni profile, which is a combination of stellar absorption and emission, in combination with further absorption by outflowing material in the ISM, we defer analysis of this line to the next section. 

\begin{figure*} [!t]
\begin{center}
\includegraphics[scale=0.4]{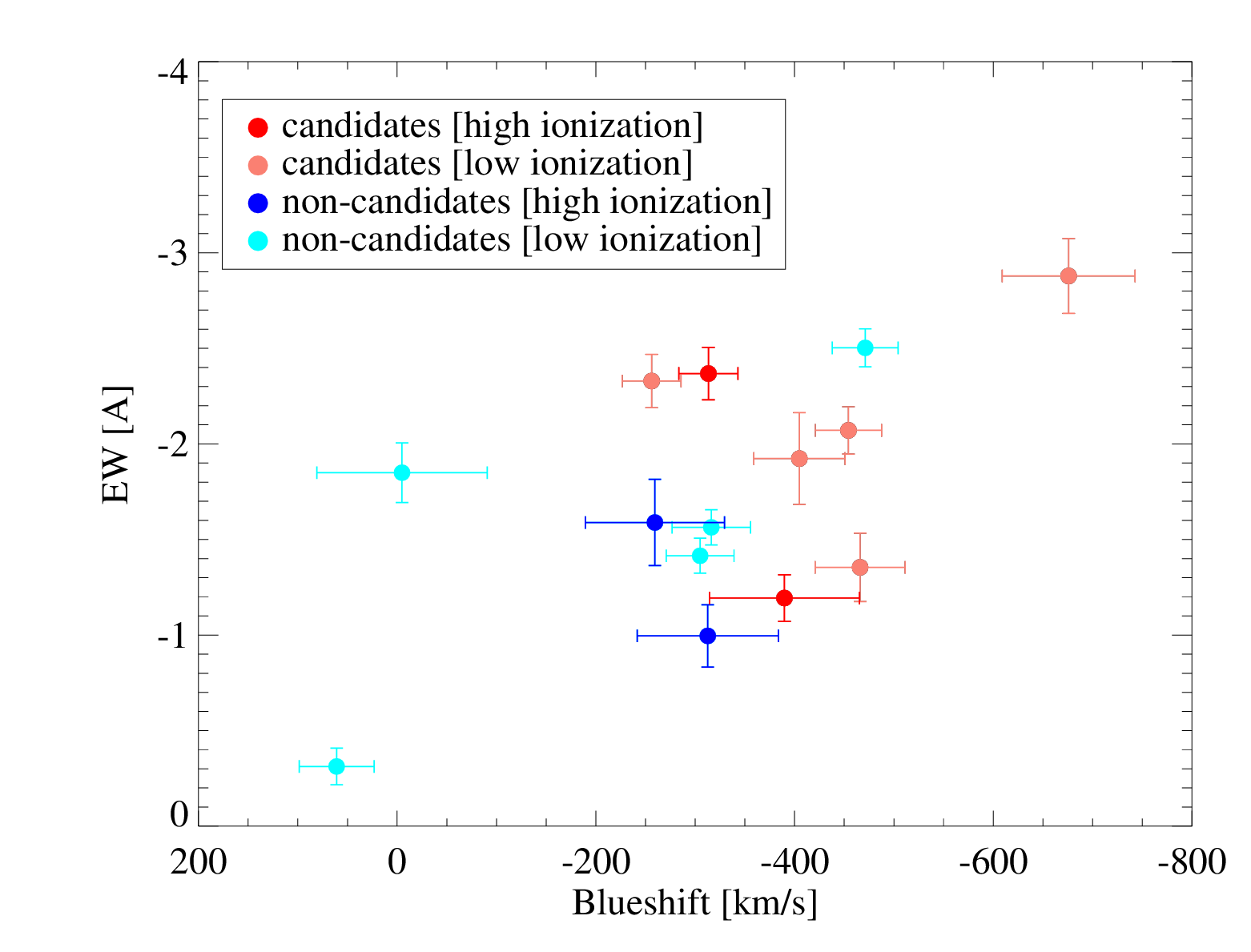}
\caption{
Equivalent width vs the blueshift of that line's centroid for all the absorption lines plotted in Figure \ref{spec_zoom}.  Lines from the candidate spectra are plotted in salmon (low ionization lines) and red (high ionization lines), and for non-candidates, cyan (low-ionization) and blue (high ionization). Candidates tend to show higher velocity blue shifts than the non-candidates, indicative that the velocity spread of the
  outflowing component is higher on average, and larger equivalent widths, indicating higher covering fraction and/or larger dispersion of velocities of absorbing interstellar clouds. }
\label{lines}
\end{center}
\end{figure*}

\begin{figure*} [!t]
\begin{center}
\includegraphics[scale=0.4]{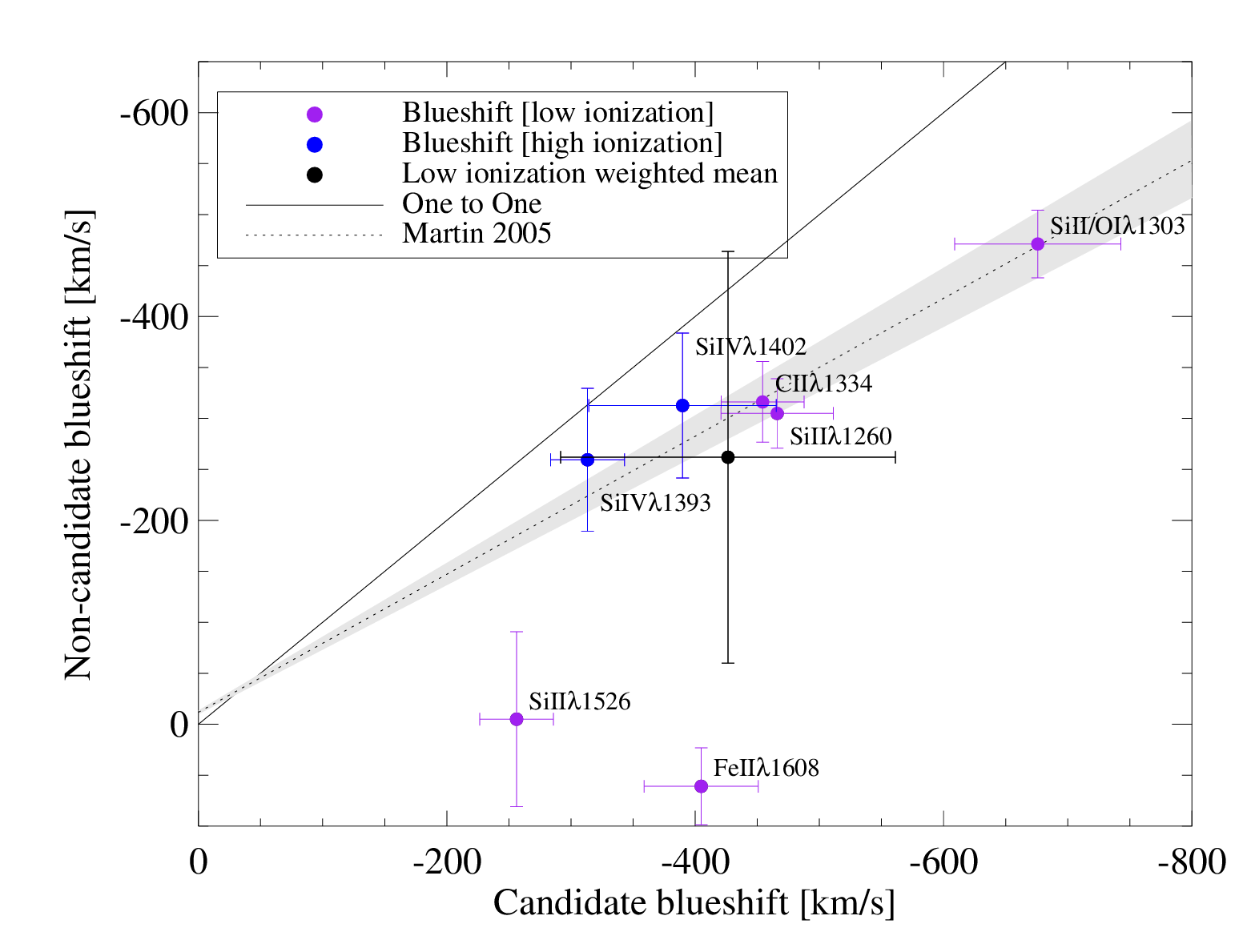}
\includegraphics[scale=0.4]{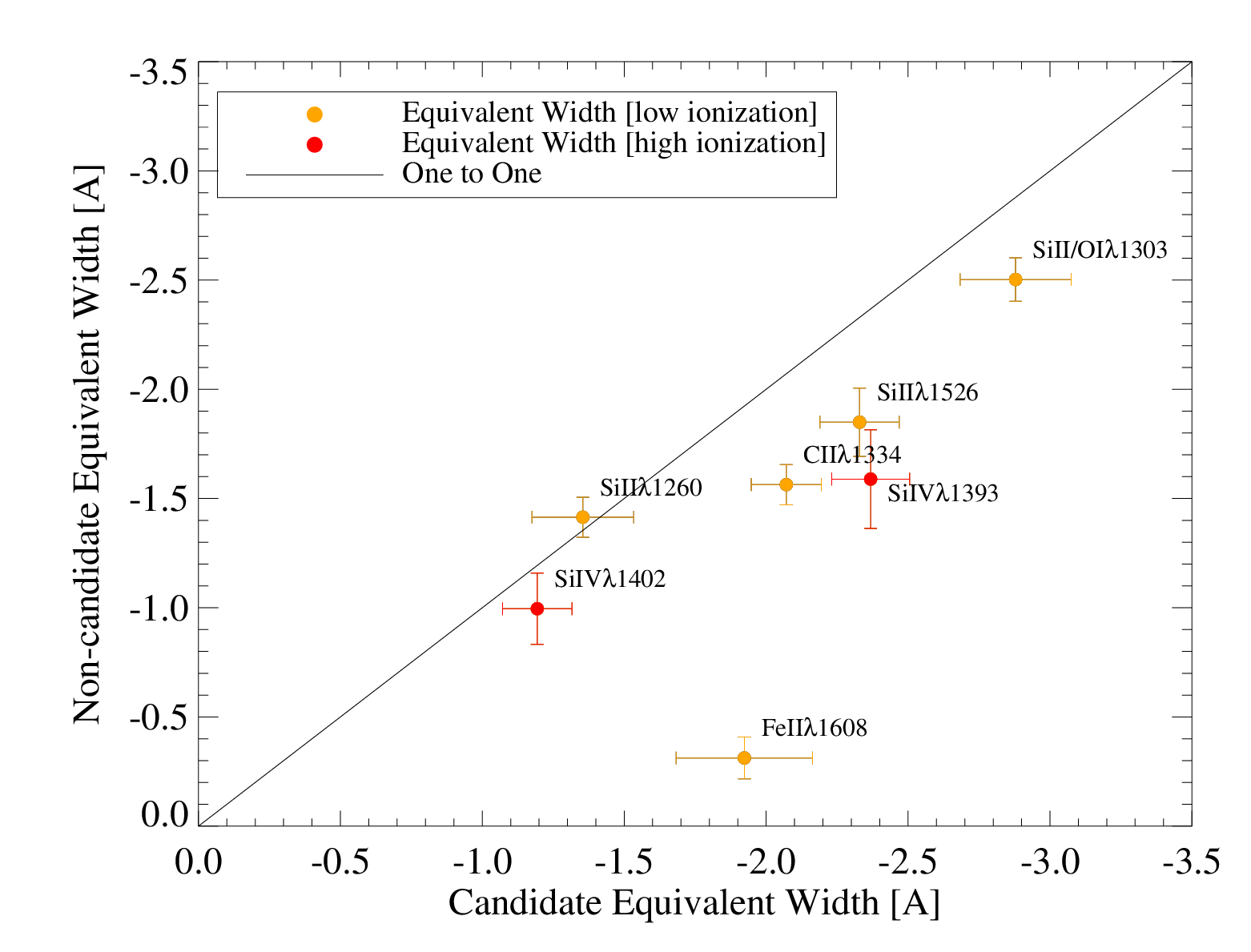}
\caption{
The same data as in Figure \ref{lines} but this time each line property measured from the non-candidate spectrum is plotted against the corresponding line property measured from the candidate spectrum. Each point represents a different absorption line. Those line properties are: blueshift of low ionization lines (purple), blueshift of high ionization lines (blue) in the left panel, equivalent width of low ionization lines (orange) and equivalent widths of high ionization lines (red) in the right panel. A point on the one to one line (black) would indicate that the measured property is the same between the two samples. Shaded region in the left panel represents the fit from \citet{Martin2005} for outflow velocity as a function of SFR.  Measurements are larger in the candidate spectrum relative to non-candidates (i.e. lines are more blue shifted, or higher equivalent width). }
\label{lines2}
\end{center}
\end{figure*}

We fit gaussian profiles to the seven interstellar lines mentioned above, where the line center, width, continuum level and peak flux density are left as free parameters. The observed and fitted absorption profiles are shown in Figure \ref{spec_zoom}. (For comparison, the same fits using the high-mass only non-candidate stack is shown in Figure \ref{spec_zoom_hm}). We also calculate equivalent widths of each line by directly integrating the line flux in the observed stack. 
 For each ISM line, we fit a line to two continuum regions bracketing the ISM line that are devoid of other features, interpolate this continuum fit at the ISM line center, and then integrate the equivalent width with respect to the continuum.
We estimate the error on the equivalent width using the standard deviation of 1000 Monte Carlo simulations generated from simulated stacks produced from gaussian variates of each point according to the rms values in the continuum regions around each line, as measured by IRAF.

We compare the line centroid relative to the rest wavelength of the line, and the line equivalent width in Figure \ref{lines}. This figure shows the equivalent width vs blueshift of
each individual line fit in Figure \ref{spec_zoom} 
for both candidates and non-candidates.
 In the figure we show low and high ionization lines color coded separately, although we do not see that there is any trend with ionization state. It is apparent however that there is a tendency for the absorption lines from the candidate stack to exhibit a stronger blueshift with respect to the rest wavelength than the non-candidates. 
 
 We also plot in  Figure \ref{lines2} the same measurements with a different representation. The left panel shows the centroid blueshift of each individual line compared directly between the candidate and non-candidate stack. In other words, the blue shift of any given line (i.e. \ion{Si}{2}$\lambda$1260) in the candidate stack is compared directly with the blueshift of that same line in the non-candidate stack. Similarly we compare directly the absolute values of equivalent width in absorption of any given line in the right panel of Figure \ref{lines2}.  The one-to-one line would indicate the exact same centroid blueshift, or equivalent width, was
  measured in candidates and non-candidates for any given absorption line. This figure shows more clearly that in nearly all measurements of blueshift and equivalent width, the candidates show a larger measurement than the non-candidates, indicating that on a line by line basis, as well as globally, candidates exhibit larger centroid blueshifts and stronger equivalent widths than non-candidates.

 This is indicative that the kinematics in the ISM of these two samples of
  galaxies may be distinct from each other. First, this offset in blueshift
  could in principle be an indication that the velocity spread of the
  outflowing component is higher on average among the candidate spectroscopic
  sample than that of the non-candidates, with more high-velocity gas
  contributing to the growth of the absorption lines in the former than in the
  latter. Another way to state this is that the smaller blueshift of the
  centroid of the non-candidates could be the result of a larger component of
  zero-velocity gas component, i.e. their ISM is on average less energized
  than that of the candidate galaxies. 
   The right panel of Figure \ref{lines2} compares the equivalent width of each absorption line from the candidate stack with that same line in the non-candidate stack. 
 These lines are saturated \citep[e.g.][]{Pettini2002, Shapley2003}, and therefore any increase in equivalent width is only weakly related to column density of that gas in the ISM.  Rather, the measured equivalent width is instead a reflection of the combined effect of the spread in velocities of the absorbing gas clouds in the ISM, and the covering fraction of gas. With this low-resolution data, it is not possible to distinguish these two contributions. The data presented  in Figure \ref{lines2}  are consistent with a scenario where the outflowing gas of the candidates  includes faster
  components and also a larger spread of velocities of its gas clouds in the
  ISM, compared with non-candidates. They also are consistent, however, with a
  lower covering fraction of the ISM in the non-candidates relative to the
  candidates, which is a distinct possibility since the morphology of the
  stellar component of the former is much more extended than that of the
  latter. It is not possible to clearly disentangle these two contribution
  with our data. As we discuss later in Section 3.3, however, the fact that the candidates
  have much stronger \Lya\ emission, given the strength of their interstellar
  absorption, compared to the non-candidates suggests that the gas covering
  fraction is unlikely playing a key role here, since a larger covering
  fraction would generally tend to depress the strength of \Lya\ emission, not
  enhance it.

\begin{table*}[!!t]
\begin{center}
\caption{Line centroid measurements\label{table2}}
\begin{tabular}{llllll}
\tableline\tableline
 Line & Rest Wavelength & Candidate blueshift [km/s] & Non-candidate blueshift  [km/s] & \\
\tableline
\ion{Si}{2}\ & 1260.42 & 465 $\pm$ 45 & 304 $\pm$ 34 \\ 
 \ion{Si}{2}\/\ion{O}{1}\ & 1303.27 & 675 $\pm$ 66 & 471 $\pm$ 33\\ 
 \ion{C}{2}\ & 1334.50 & 454 $\pm$ 33 & 316 $\pm$ 39\\ 
 \ion{Si}{4}\ & 1393.76 & 313 $\pm$ 29 & 259 $\pm$ 70\\ 
 \ion{Si}{4}\ & 1402.77 & 389 $\pm$ 75 & 312 $\pm$ 71\\ 
 \ion{Si}{2}\ & 1526.71 & 256 $\pm$ 29 & 4 $\pm$ 85\\ 
 \ion{Fe}{2}\ & 1608.45 & 404 $\pm$ 45 & -60 $\pm$ 38\\ 
\tableline
\end{tabular}
\end{center}
\end{table*}

Since the candidates on average have higher SFRs than non-candidates, we investigate how these velocities relate to each other in terms of the differences in SFR. Relations between outflow velocity and SFR have been observed over many orders of magnitude in galaxy SFR, and \citet{Martin2005} identified a relation between outflow velocity and SFR, such that V$\propto$SFR$^{0.35}$. 

This relation is based on the upper envelope of the distribution of velocity vs SFR for galaxies at z$\sim$0 over 4 orders of magnitude in SFR (the full distribution studied there exhibits a lot of scatter on an individual galaxy basis).

Other studies have found similar relationships between SFR and outflow velocity for a variety of galaxy populations \citep{Weiner2009, Banerji2011, Song2014}, (although still other studies have found that a more significant relationship exists between outflow velocity and surface density of star-formation, see a more complete discussion in Section 4.2). Here, we use the observed relation from \citet{Martin2005} to see if our observed average velocity differences and SFRs are consistent with this relation. The relation is given by log$V=$(0.35$\pm$0.06)log(SFR)$+$(1.56$\pm$0.13), where V is outflow velocity in \kms and SFR is in M$_{\odot}$yr$^{-1}$. We plot this relation in Figure \ref{lines} (dashed line), where we use the relation to predict the ratio of the average outflow velocities, given the average SFRs of the two samples. The grey region outlines the region within the uncertainties of the fit parameters for this relation \citep{Martin2005}. We see that the average low-ionization line blueshift of the two stacks follow this empirical scaling relation between SFR and outflow velocity for galaxies at lower redshifts.  The weighted averages and standard deviations of the low-ionization lines, which generally trace the same gas and therefore may have similar kinematics, are also presented. Candidate low-ionization lines exhibit on average centroid velocities and standard deviations of -426$\pm$134 km s$^{-1}$, and that of non-candidates is -262$\pm$202 km s$^{-1}$. \citet{Martin2005} notes that although the upper envelope of the velocity-SFR distribution for the lower redshift sample of galaxies studied there follow this relation, at SFR $>$ 10 the scatter in velocity at a given SFR is huge, and the slope may in fact be shallower. If such a scaling relation exists, and is flatter than that suggested by \citet{Martin2005} and \citet{Weiner2009}, the difference in centroid velocities between our two samples may be more significant than that expected purely from SFR differences using the scaling relation of the upper envelope. Regardless, we find that these velocities are consistent with what is expected given the differences in their SFRs.

These blueshift measurements obviously rely on a robust measurement of systemic redshift in each galaxy, and as discussed previously, 
the systemic redshift we measure is subject to some uncertainty. With these data this is unavoidable.
  However, it is possible to get a first order check on our conclusion of the difference in {\it relative centroid} velocities between the two samples by comparing the net offsets in velocity between the absorption lines and the Ly$\alpha$. Because Ly$\alpha$ is a resonant line, it is generally seen redshifted with respect to systemic in high-redshift galaxies, because Ly$\alpha$ photons which scatter off outflows for example allows the line to redshift out of resonance \citep{Shapley2003, Steidel2011}. Indeed, LBGs generically show a difference in velocity between Ly$\alpha$ and interstellar absorption lines that is smaller than the difference between rest wavelengths, suggesting Ly$\alpha$ scatters off outflows in the background of galaxies (i.e. moving away) and interstellar lines absorbed by outflowing gas in the foreground (i.e moving towards us). 
   
\begin{figure} [!!t]
\begin{center}
\includegraphics[trim=2cm 0cm 0cm 0cm,scale=.3]{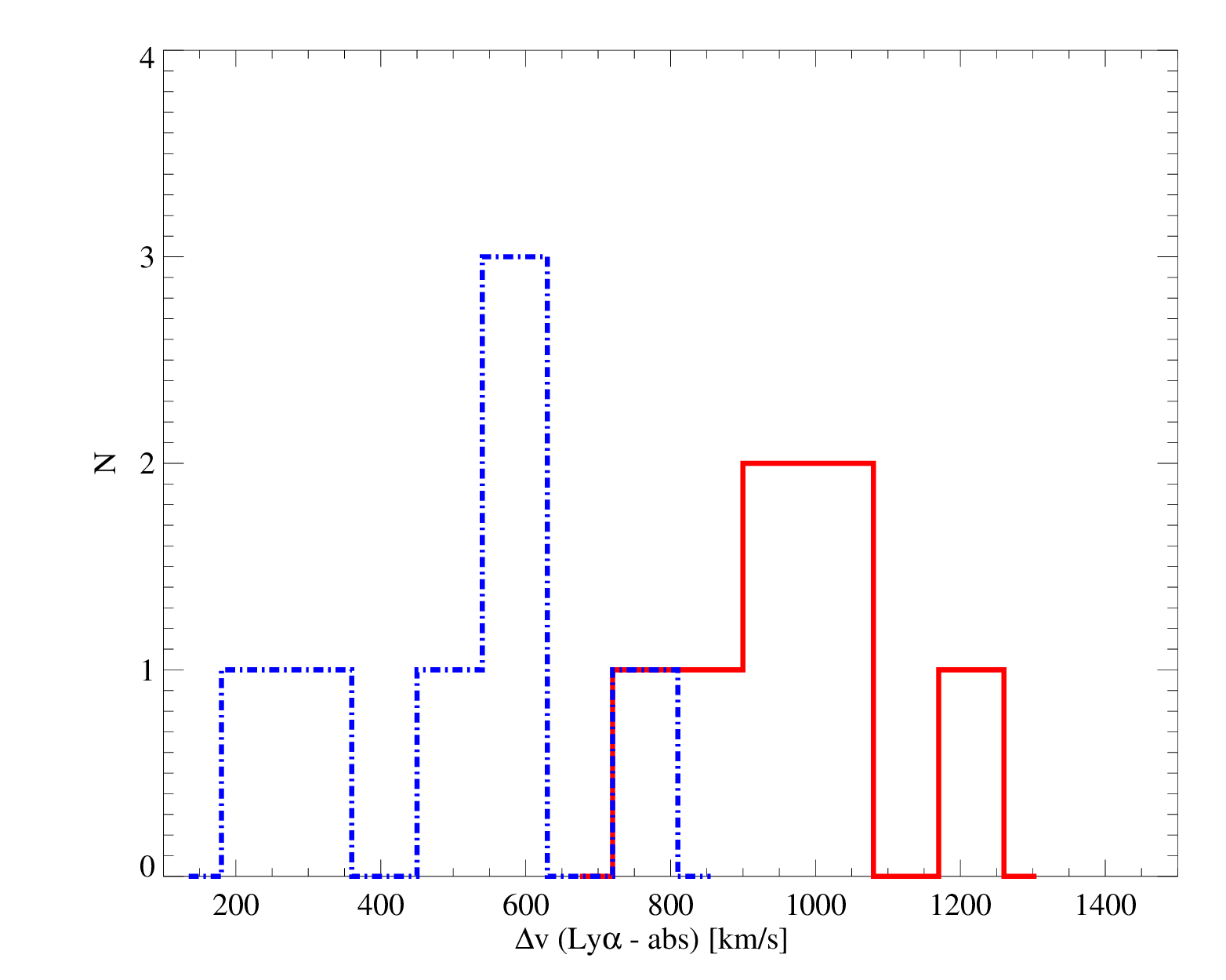}
\caption{Histograms showing the velocity offset between \Lya\ and each absorption line presented in Figure \ref{lines}. Candidates (red) show larger  velocity offsets than non-candidates (blue).}
\label{veloff}
\end{center}
\end{figure}

  As can be seen in Figure \ref{lya}, both samples exhibit Ly$\alpha$ profiles which are redshifted, as would be expected if it traces outflowing gas and must be redshifted out of resonance in order to travel freely,  or, if it is absorbed by neutral gas close to rest with respect to the galaxy. It is clear that
the candidate Ly$\alpha$ profile exhibits a larger redshift, whereas the non-candidate Ly$\alpha$ profile is closer to the rest wavelength. In Figure \ref{veloff}, we plot a histogram of the velocity difference between the peak wavelength of the Ly$\alpha$ profile, and the Gaussian centroid of the interstellar absorption lines. 
 This figure shows that the net offsets in velocity differ significantly between the two samples, with the candidates showing larger relative velocities between interstellar absorption and Ly$\alpha$ emission, independent of systemic velocities.

\begin{figure*} [!t]
\begin{center}
\includegraphics[trim=2cm 0cm 0cm 0cm,width=4in]{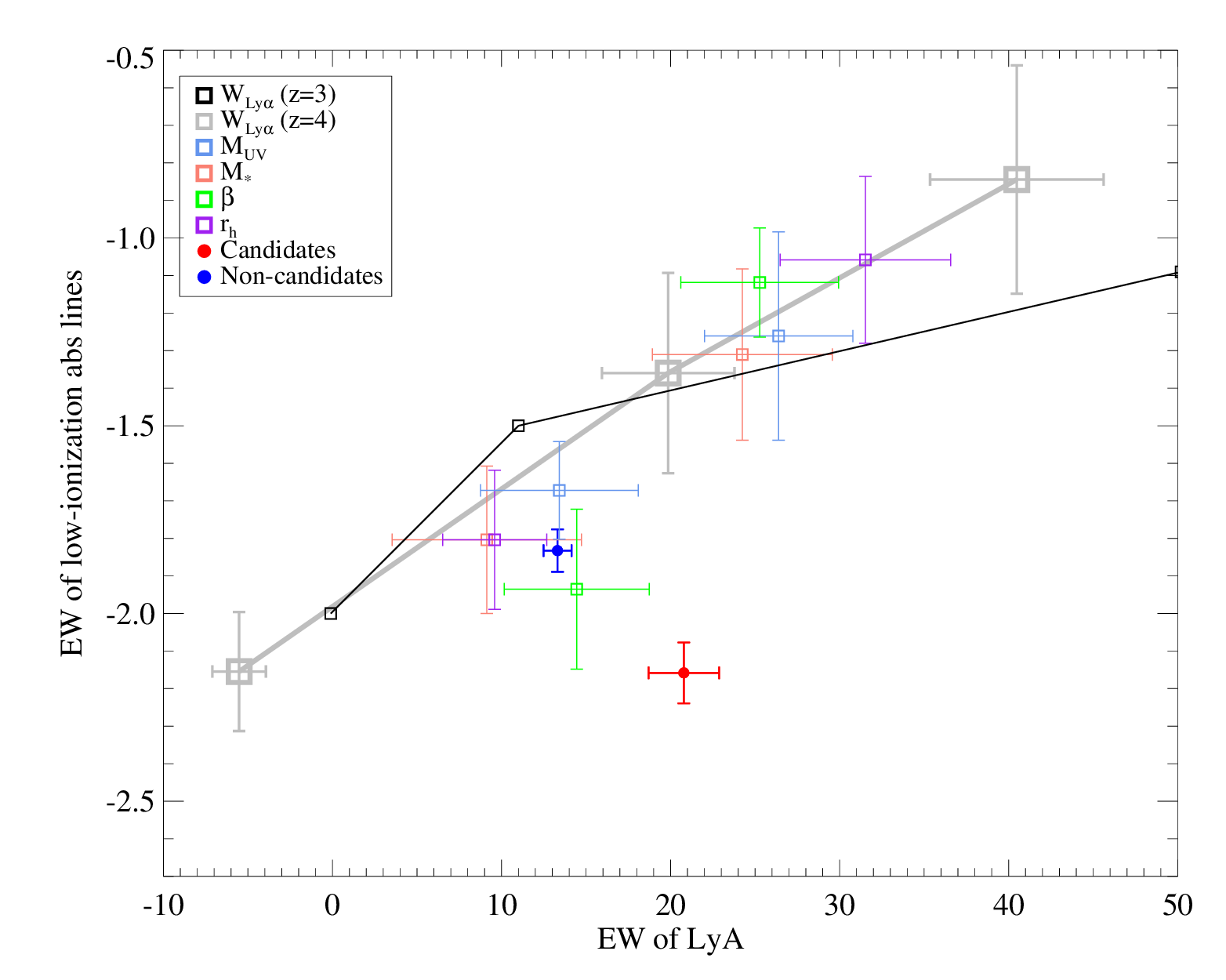}
\caption{Comparison of \Lya\ equivalent width with that of the average equivalent width of low-ionization absorption lines in the ISM. Low-ionization ISM lines used to define the average equivalent width are \ion{Si}{2}$\lambda$1260, OI/SiII$\lambda$1303, \ion{C}{2}$\lambda$1334, and \ion{Si}{2}$\lambda$1526. The data are from \citet[][black squares]{Shapley2003}, and \citet[][grey and colored squares]{Jones2012}, where these points are binned by properties in the legend (see text for full description). While non-candidates display properties consistent with the trend of normal LBGs, candidates lie significantly off the trend, simultaneously displaying strong \Lya\ and strong interstellar absorption. Figure adapted from \citet{Jones2012}}
\label{ewew}
\end{center}
\end{figure*}

\subsection{Ly$\alpha$ properties}

As is clear from the large jackknifed error bars on the stacks in Figure \ref{lya}, which reflect
the variation among  the individual spectra, the LBGs making up both our samples
exhibit a variety of \Lya\ properties. In the candidate stack, three galaxies
lack \Lya\ coverage, and of the remaining, about half exhibit \Lya\ in
emission (one weak detection, two with detected \Lya\ in excess of the average
and two with detected \Lya\ below the average). Of the non-candidates, one
galaxy lacks \Lya\ coverage, and 9 show well detected \Lya\ emission, 4 have
weak detections, and 6 are not detected. To summarize, roughly half of each
sample with \Lya\ spectral coverage has \Lya\ in emission and the other half
in absorption. This is in good agreement with the global statistics observed
among LBG at $z\sim 3$ in general, in which roughly half have \Lya\ detected
in emission \citep[e.g.][]{Shapley2003}. Thus, although the two samples of spectra that
enter in both stacks are relatively small, we do not believe that
small--number statistics are significantly affecting the average properties
of each LBG sample as represented by the two stacks.  We further note that removing the 4 galaxies without \Lya\ coverage from the stacks does not change our results.

Since the SFR and overall UV spectra of the two samples are similar, the
differences between their Ly$\alpha$ lines very likely reflect differences in
the coverage and conditions of neutral Hydrogen in the ISM. As discussed
previously, we observe the candidate Ly$\alpha$ profile to be more redshifted
than that of the non-candidates. 
 However, the presence of
Ly$\alpha$ emission blue shifted with respect to the rest wavelength may
indicate the escape of Ly$\alpha$ photons through patchy or ionized foreground
ISM \citep[e.g.][]{Heckman2011} or be an indication of infall
\citep[e.g.][]{Kulas2012}. In this section, we describe our measurement of the
equivalent width of the Ly$\alpha$ line, and present analysis based on this
measurement.

In addition to being characterized by redshifted emission, the Ly$\alpha$ line
in our samples, like many galaxies at similar high-redshift, is characterized
by asymmetric profiles, including some blue shifted emission, as well as blue
shifted absorption. Measurement of the intrinsic equivalent width of
Ly$\alpha$ at high-redshift is complicated by additional intervening
absorption due to neutral Hydrogen in the circum--galactic medium of the
galaxy and by the Ly$\alpha$ forest. This extrinsic absorption, i.e. not
intrinsic to the galaxy itself, makes defining the blue ward edge and shape of
the Ly$\alpha$ profile highly uncertain. To ensure an equal comparison with
previous measures in the literature, we adopt a similar procedure to that by
\citet{Shapley2003, Stark2010, Jones2012} when we measure the equivalent width
of the \Lya. Using the stacked spectrum of z$\sim$3 of \citet{Shapley2003}, we
have reproduced their measurement of Ly$\alpha$ equivalent width of 14.3\AA\
using a continuum level defined by the mean value between rest wavelengths
1225-1235\AA, and where the bounds for the line are defined as follows. The red
ward edge of the Ly$\alpha$ profiles is defined by where the flux in the line
meets the continuum level, and the blue ward edge is defined to be as far from
the rest wavelength of Ly$\alpha$ as the point where the red side of the
profile hits the continuum, i.e. we use for the blue side of the line the same
half--width at zero level observed for the red side. We found that this
procedure reproduced the documented equivalent width in \citet{Shapley2003}
for their LBG stack within our error on the equivalent width. We estimate our
error on the equivalent width as described in Section \ref{kinprop}, except we
use the jackknifed error on the stack to produce gaussian variates on the
stack, since this error is much larger than the continuum rms in the vicinity
of Ly$\alpha$.  Following this procedure, we find that the candidate stack is
characterized by much stronger Ly$\alpha$ equivalent width (20.8$\pm$2.1{\AA})
than the non-candidate stack (13.3$\pm$0.8{\AA}). Furthermore, we note that
non-candidates exhibit a larger equivalent width of blue shifted \Lya\ than
the candidates, and makes up a higher fraction of the total equivalent width
in \Lya\ (about 15\%, compared to 7\% in the candidate stack).

We compare our measurement of \Lya\ equivalent widths to the general trend of
equivalent widths in \Lya\ and low-ionization interstellar absorption lines
found among z$>$3 LBGs in Figure \ref{ewew}, which is adapted from
\citet{Jones2012}. The figure shows that,
in general, the \Lya\ equivalent width correlates with the equivalent width of
low-ionization interstellar absorption lines. The sense of the correlation is
that the larger the equivalent width of low-ionization interstellar absorption
lines the smaller the emission equivalent width of Ly$\alpha$, or the larger
the absorption equivalent width of
\Lya\ \citep{Shapley2003,Vanzella2009,Jones2012, Berry2012}.  This is
generally interpreted to indicate that \Lya\ is present in galaxies with
patchy neutral covering fraction of gas in the ISM, or clumpy ISM
\citep[e.g.][]{Shapley2003, Quider2009, Jones2012, Jones2013}, since for Ly$\alpha$ to
escape the galaxy it must travel freely without neutral hydrogen in the
foreground. Alternatively, \Lya\ can escape if it is redshifted out of
resonance \citep[e.g.][]{Verhamme2008,Steidel2010}.

Figure \ref{ewew} also shows the general relation between the equivalent
widths of the low-ionization ISM lines and  \Lya\ 
for z$\sim$4 LBGs that have been split into bins according
to each of the following properties; equivalent width of
\Lya\ ($W_{Ly\alpha}$), UV luminosity (M$_{UV}$), stellar mass (M*), UV slope
($\beta$), and SExtractor half-light radius (r$_{h}$). The cluster of colored
points at low \Lya\ equivalent width are the half of z$\sim$4 galaxies which
are each the brightest, more massive, redder UV slope and larger
radii. \citet{Jones2012} note that all trends in Figure \ref{ewew} with
galaxy properties listed above are due to their correlations with either the neutral gas covering fraction, or the kinematics in the ISM.

The non-candidate stack has interstellar absorption and \Lya\ emission consistent with the general trend of z$\sim$3 LBGs.  However the candidate stack does not, and instead has stronger \Lya\ emission than expected given the large (negative EW) absorption.

Differences in morphology appear to correlate with this trend, as can be seen
by the purple points in this figure. When split by half-light radius, larger
galaxies make up the low equivalent width of \Lya\ part of the trend, while
small galaxies have stronger Ly$\alpha$. \citet{Vanzella2009, Law2012b} find a
relationship between Ly$\alpha$ and half-light radius, such that smaller
galaxies are more likely to have Ly$\alpha$ in emission \citep[see also][]{Malhotra2012}, although \citet{Pentericci2010} find no dependence of Ly$\alpha$ properties and UV
morphology. Our two samples are not explicitly split in size, they are roughly
split in the mass-size diagram (see Figure \ref{sampleprop}) and therefore
comparison in the context of the trend with galaxy size may be complicated by the
overlap in their size distributions. However, if we compare the candidates to
galaxies in the small size bin, the discrepancy between the candidates and the
LBG trend of \Lya\ and low-ionization lines becomes even larger. Given their
smaller size on average, it might be expected that candidates should exhibit
weaker interstellar absorption than non-candidates.  The candidates, however, are not just smaller, they are
denser and have a generally larger stellar mass and star formation rate such
that, if they quench soon after $z\sim 3$, they are passively evolving by
$z\sim 2$ and have stellar mass $M>10^{10}$ \msun. We will discuss plausible
interpretations of this strong deviation among candidates in Section
\ref{Discussion}.

\begin{table}[!!t]
\begin{center}
\caption{Equivalent Width Measures\label{table1}}
\begin{tabular}{llllll}
\tableline\tableline
Galaxy Sample & Line & equivalent width [{\AA}]&  & \\
\tableline
Candidates &\ion{C}{4}\  [abs] & -5.2 $\pm$ 0.3 & \\ 
 &\ion{C}{4}\  [em] & 0.9$\pm$ 0.1 &  & \\
 & \ion{He}{2}\  & 1.9 $\pm$ 0.2 &  & \\
& \Lya\ & 20.8 $\pm$ 2.1 & & \\

Non-candidates &\ion{C}{4}\  [abs] & -3.0 $\pm$ 0.2 & & \\  &\ion{C}{4}\  [em] &  0.3 $\pm$ 0.1 &  & \\
  & \ion{He}{2}\  &  Undetected [rms=0.09] &  & \\
& \Lya\ & 13.3 $\pm$ 0.8 & & \\

\tableline
\end{tabular}

\end{center}
\end{table}

\subsection{X-ray properties}

In this section, we investigate any evidence in the data that either sample is
affected by AGN activity. In \citet{Williams2014}, the X-ray and infrared
properties of the parent samples of candidate and non-candidate LBGs were
studied. Only 3 non-candidates were detected in the Chandra 4 Ms data, and
X-ray stacking indicated no statistical difference between the two
samples. Additionally stacks in the Spitzer 24$\mu$m and Herschel 100$\mu$m
images indicated no significant average IR emission among candidates, and a
minor (4$\sigma$) 24$\mu$m detection among non-candidates, i.e. possible
evidence of marginally larger dust obscuration. For the spectroscopic
subsamples studied here, no galaxies are X-ray detected, 3 galaxies are
detected in the 24$\mu$m images (2 candidates, 1 non-candidate) and one of
those two candidates has a 100$\mu$m detection based on the 24$\mu$m
position. The high number of non-detections may very well be due to the fact
that fluxes at these wavelengths for z$>$3 galaxies are below the detection
limits of the survey, rather than indicating a lack of (faint) AGN
presence. To test for this possibility we have repeated the X-ray stacking in
the Chandra 4 Ms imaging for this subsample of LBGs as in
\citet{Williams2014}.  We find that the spectroscopic candidates show a very
marginal detection in the soft band only, with only 1.6$\sigma$
significance. The average luminosity implied is 3.8x10$^{41}$ erg
sec$^{-1}$cm$^{-2}$, too low to be the result of AGN activity. There is no
significant detection in the hard band, suggesting that the soft--band
detection, if real, is most likely due to star formation rather than AGN
activity.  The non-candidate sample do not show significant stacked signal in
either band.

\subsection{\ion{C}{4}\ and \ion{He}{2}\ Properties}

\begin{figure*} [!t]
\begin{center}
\includegraphics[trim=2cm 0cm 0cm 0cm,width=5in]{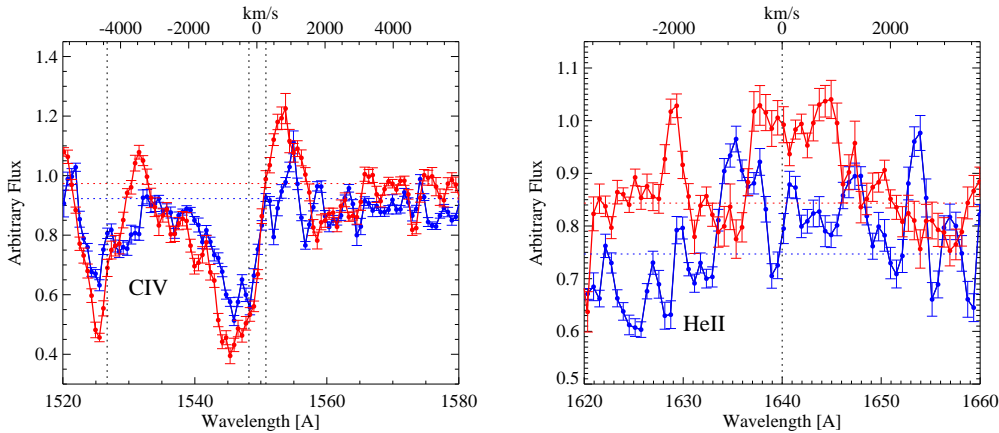}
\caption{Zoom-ins of\ion{C}{4}\  and \ion{He}{2}\  visible in candidate (red) and non-candidate (blue) spectra in Figure \ref{spec}.  Errorbars reflect sample standard deviation, as described in section \ref{stackproc}. The horizontal dotted lines indicate the continuum level measured from the candidate (red) and non-candidate (blue) stacks for each line.}
\label{civ_heii}
\end{center}
\end{figure*}

Figure \ref{civ_heii} shows that candidates have larger equivalent width of
\ion{C}{4}\, both in the absorption and emission components of the P-Cygni profile, as well as \ion{He}{2}\ in emission. The non-candidate spectrum however, exhibit smaller equivalent widths in both absorption and emission components of \ion{C}{4}\, and have no detected \ion{He}{2}\ in emission. We measure the equivalent widths as in Section 3.2, and list the measured equivalent widths for each sample in in Table \ref{table1}. In this section, we present analysis of AGN contribution and constraints on the properties of their stellar populations based on these two features in the spectra of the two samples.

\subsubsection{Emission lines and AGN}
The presence of AGN can be identified by broad emission lines, and
assessed using a host of emission line diagnostics
\citep{BaldwinPT1981,VeilleuxOsterbrock1987,DopitaSutherland1996,Kewley2006}. Unfortunately without rest frame optical
(observed near-infrared) spectroscopy we miss a large set of potential
emission line probes of AGN.
Rest frame UV spectroscopy of narrow-line and broad-line AGN have been studied
among the LBG samples in \citet{Steidel2002, Hainline2011}. The stacked
spectra of these LBGs with AGN differ significantly from either of our
samples, in that they exhibit strong and broad emission in many high
ionization lines including \ion{N}{5}\, \ion{N}{4}\ ], and \ion{C}{4}\, and also
the low ionization line \ion{He}{2}\ . We detect neither \ion{N}{5}\ nor \ion{N}{4}\ ].
We detect both \ion{C}{4}\ and
\ion{He}{2}\ in emission in the candidate spectrum, but their properties are
very different from what is observed in AGN. The \ion{C}{4}\ emission is
clearly a P-Cygni profile and, like \ion{He}{2}\, is commonly observed among
star--forming galaxies and lacks the strong and broad emission characteristic
of AGN. We compute the ratio of the equivalent widths of \ion{C}{4}\ and
\ion{He}{2}\  to compare with typical AGN values.

We find an equivalent width ratio of \ion{C}{4}\ to \ion{He}{2}\ of 
0.5$\pm$0.1,  
well below the AGN diagnostic ratio $>$ 1.5 for narrow-line
AGN and high-redshift radio galaxies \citep{Mccarthy1993, CorbinBoronson1996,
  Humphrey2008, Matsuoka2009},
showing that even if an AGN were present, it would be very weak and in any
case not a significant contributor to the emission lines.

As a comparison,
\citet{Hainline2011} find that the equivalent widths of \ion{C}{4}\ and
\ion{He}{2}\ in a composite spectrum of narrow-line AGN in LBGs at
z$\sim$2-3 would yield a line ratio of $\sim$2$\pm$0.2, well in excess of what
we find in the candidate stack. Their subsample of AGN which have Ly$\alpha$
equivalent width $<$ 63{\AA}, which is their subsample of AGN with the weakest
emission lines and the lowest \ion{C}{4}/\ion{He}{2}\ ratio, is still in
excess of what we find, with a ratio equal to 1.5$\pm$0.4). Although this may
not be a conclusive rejection of AGN presence in these galaxies, to the best
of the ability of the data we have available, we do not see obvious evidence
 that either sample is dominated by AGN activity. 
 Further exploration of this
possibility will be possible with NIR spectroscopy, to access the rest-frame
optical emission lines in these galaxies.

\subsubsection{Contribution from massive stars and their properties}

Both the \ion{C}{4}\ P-Cygni feature and the \ion{He}{2}\  emission line, which originate from
WR and other massive stars, are specifically related to the stellar
metallicity, because the higher opacity results in increased radiation pressure
\citep[][]{VinkdeKoter2005}.  Additionally, the mass loss rates for WR
stars appear to be metallicity dependent, with higher mass loss rates for increased
metallicity \citep[][]{VinkdeKoter2005}. Thus, the ratio of WR stars
to O type stars correlates with metallicity, in the sense that at higher
metallicity, more O stars evolve through a WR phase because the mass threshold
for evolving through the WR phase is lower. At lower metallicity the
threshold for a WR phase is at a much higher mass, and therefore, for a given
initial mass function (IMF), fewer stars will become WR \citep{Maeder1991,
  Meynet1995, crowther2002, GrafenerHamman2005, MeynetMaeder2005,
  LopezSanchez2010}. Therefore, an increase in the metallicity of a galaxy
increases the likelihood of observing WR-associated features
\citep{Brinchmann2008, Brinchmann2008b}.

The presence and luminosity of the \ion{He}{2}\ emission line in the spectra of WR stars also depends on the metallicity \citep{CrowtherHadfield2006}, and in low-metallicity WR stars, the line may be absent altogether. 
 \citet{Brinchmann2008} find that in an
(instantaneous) starburst model, the \ion{He}{2}\  emission from O-type stars will
increase until the emission peaks during the WR phase of the stars formed in
the burst. In the case of continuous star-formation, however, the equivalent
width of \ion{He}{2}\  will eventually plateau at equivalent width $\sim $1.5\AA until star-formation
ceases, provided the metallicity is at least half solar
\citep{Brinchmann2008}. It is interesting to note that in Table \ref{table1}
we measure an equivalent width in \ion{He}{2}\ of 1.9$\pm$0.3{\AA}, roughly consistent with
their value. In any case, the presence of the line in the candidate average
spectrum and its absence in that of the non-candidates suggests that the
former have higher metallicity than the latter, in agreement with the relative
strength of the \ion{C}{4}\ feature.

\begin{figure*}[!!t]
\begin{center}
\includegraphics[trim=7cm 0cm 0cm 2cm,scale=0.35]{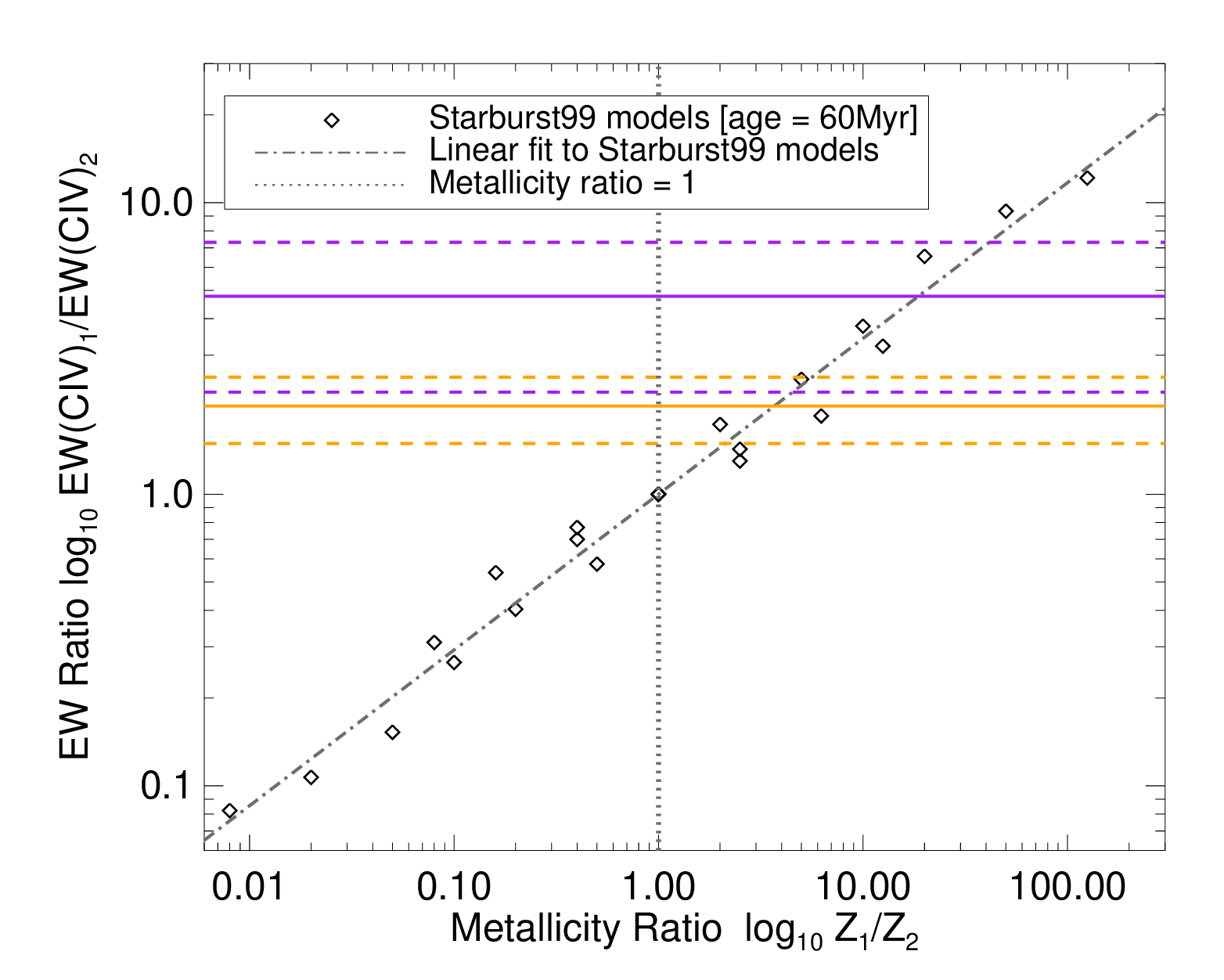}
\includegraphics[trim=0cm 0cm 7cm 2cm,scale=0.35]{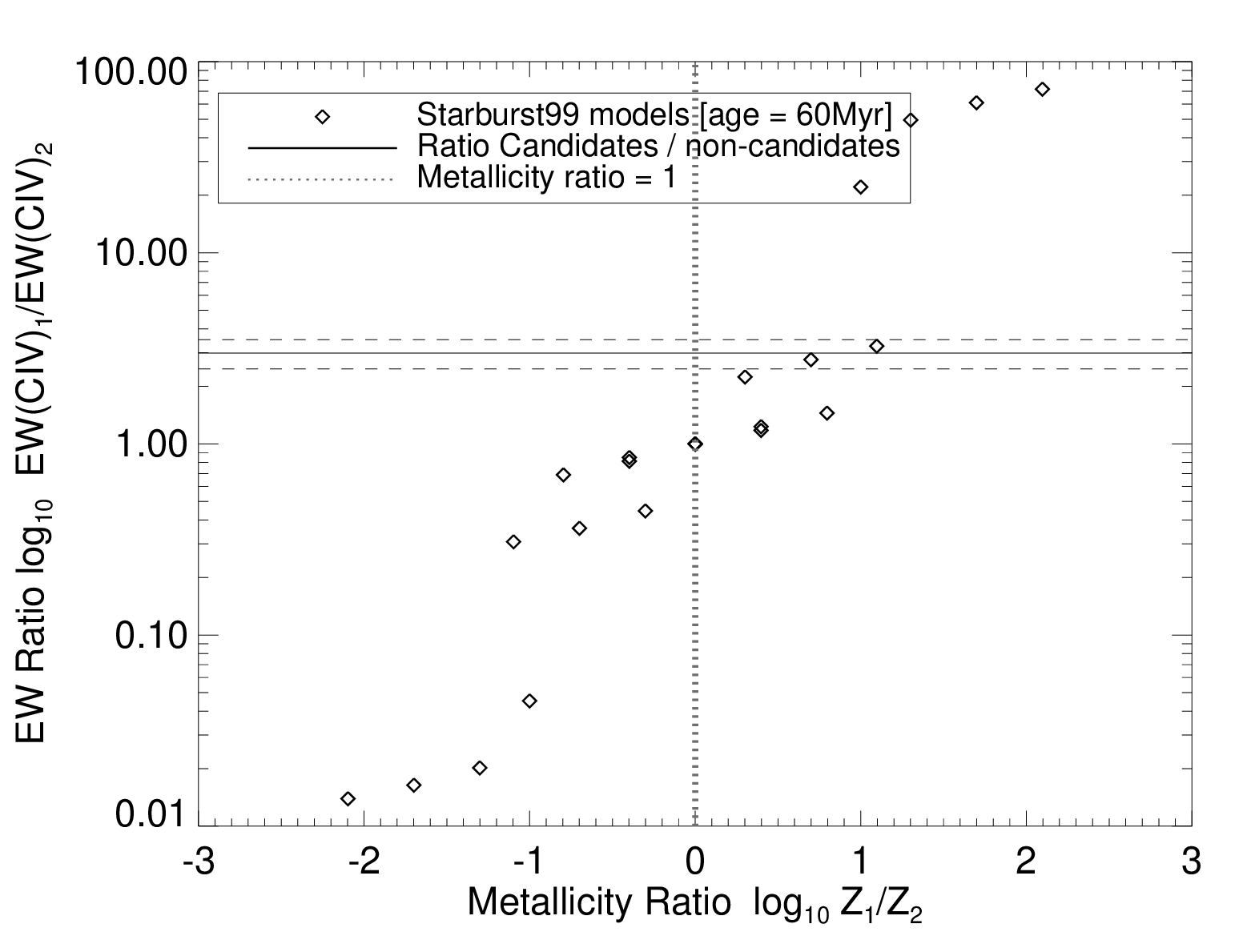}
\caption{Metallicity is related to equivalent width (EW) of \ion{C}{4}, such that equivalent width increases with increasing metallicity. The relationship between EW of \ion{C}{4} and metallicity is quantified using Starburst99 models (Leitherer et al. 2010) of varying metallicity generated as described in the text. Left panel: The relationship between the ratio of equivalent width in \ion{C}{4} (absorption component) for any two Starburst99 models as a function of the ratio of the metallicities of those models (points). 
 Each diamond point represents the ratio of equivalent width in \ion{C}{4}\  absorption of two models, as a function of the metallicity of ratio. The observed ratio in \ion{C}{4}\  equivalent width of candidates to non-candidates (stellar absorption component, with approximate ISM contribution removed), is indicated by the horizontal lines (purple lines are the ratio where the ISM contribution is taken from the \ion{Si}{4}$\lambda$1393 line, and orange lines are the more conservative estimate of the ISM contribution from the line with the largest equivalent width. Dotted lines are propagated equivalent width errors. Right Panel: Same as left panel, but the equivalent width of \ion{C}{4}\  is the emission component. The observed ratio of the emission component is the solid line, and dashed lines indicates propagated equivalent width errors. Candidates appear more metal enriched than non-candidates based on their \ion{C}{4} properties.}
\label{metrat}
\end{center}
\end{figure*}

In the rest of this section, we investigate evidence for metallicity differences using what indicators we have available in the rest frame UV. \citet{Rix2004} established a variety of UV metallicity indicators, based on photospheric blends of iron in massive stars (the 1978{\AA} index). Due to the high-redshift of our galaxies and relatively noisy spectra, we do not have good signal to noise in this spectral region of our stacks. Other metallicity indices exist based on photospheric blends between 1370-1500{\AA} \citep{Leitherer2001, Rix2004,Sommariva2012},  in regions of our stacks with higher signal-to-noise. These indices have been used as metallicity indicators using bright lensed LBGs \citep{Quider2009, Quider2010, DessaugesZavadsky2010},
and also on a stack of 75 spectra at z$\sim$2 \citep{Halliday2008}.
It is unclear if with the noisiness of the continuum in our stacks and the redshift uncertainties that these would provide robust measures of metallicity. However, as mentioned above, stellar wind lines in the UV are metallicity dependent, and \citet{Keel2004} also noted the metallicity dependence of the \ion{C}{4}\  P-Cygni profile properties in observations, very similar to those in spectral synthesis codes \citep{Rix2004, EldridgeStanway2009}, such that both the absorption and emission components increase their equivalent width with increasing metallicity. Such properties of the winds of hot massive stars, and their variation with metallicity, have been incorporated into the Starburst99 library of theoretical UV spectra \citep{Leitherer1999, Leitherer2010}. 

Although it is not possible to use the exact equivalent width to estimate the average metallicity of the galaxies in the stack using these models, we can use them in a {\it relative} way to try to estimate how much more metal rich the candidates may be than the non-candidates, based on their stellar wind signature. To illustrate further the dependence on metallicity of both the emission and absorption component of the \ion{C}{4}\  P-Cygni profile in (purely stellar, excluding interstellar absorption) spectra, we have calculated the equivalent width of both components in Starburst99 models of varying metallicity. We have generated models 
at the five metallicity intervals available (0.05, 0.2, 0.4, 1, 2 Z$_{\odot}$) with continuous star-formation
 at an age of 60 Myr. We convolve the model spectra with a gaussian of FWHM matching that of our observed VLT spectra, and calculate the equivalent width in both the emission and absorption components of \ion{C}{4}\  from each model. To illustrate the variation in \ion{C}{4}\  properties between any two models of differing metallicity, we calculate the equivalent width ratio (i.e. equivalent width of \ion{C}{4}$_{1}$/ equivalent width of \ion{C}{4}$_{2}$) and plot it as a function of metallicity ratio, Z$_{1}$/Z$_{2}$ in Figure \ref{metrat}. The left panel of Figure \ref{metrat} shows this for the stellar \ion{C}{4} absorption component from the models (points), where it is obvious that the equivalent width ratio in stellar absorption is strongly correlated with the metallicity ratio. The equivalent width ratio is related to the metallicity ratio as a well-fit power-law. The \ion{C}{4} absorption in our spectra, however, are a blend of both stellar absorption as in the Starburst99  models, but also interstellar absorption, and from these data it is impossible to separate the two components. Therefore, we cannot use the measured  \ion{C}{4} absorption as an estimate of metallicity. 
 
   The emission component, however, both in the models and our observed stacks is indeed stellar in origin. In the right panel of Figure \ref{metrat}, we repeat the same comparison for the emission component. The variation in equivalent width of the emission component is a bit more complicated than the absorption, but a trend of increasing equivalent width with metallicity is still apparent. We measure the observed ratio of equivalent width in the \ion{C}{4}\  emission in the candidate stack to that of the non-candidates, and include this as a horizontal line in the right panel of Figure \ref{metrat}. The dashed lines represent the propagated errors from the equivalent width measurements. 
Although the complicated nature of the equivalent width ratio of emission as a function of metallicity ratio precludes a robust translation of equivalent width ratio to value of metallicity ratio (as for example might be possible using the linear fit to the absorption component), the intersection of the observed ratio with the relation from the models clearly corresponds to a ratio larger than 1, indicative that the candidates have larger metallicity than non-candidates.

Although the emission components of \ion{C}{4} are stellar in origin, their equivalent width may be eroded by interstellar absorption near systemic, and it is impossible to tell if it preferentially does so in one sample vs the other. Therefore, we see if the \ion{C}{4} absorption component at least agrees with our conclusion of increased metallicity after some simple assumptions.

In both stacks, \ion{C}{4} absorption equivalent width exceeds that of the other ISM lines on average by a factor of two, indicative of the additional contribution of stellar absorption. To compare the measured absorption component of \ion{C}{4} equivalent width from the stacks to the Starburst99 models, we first use the other ISM absorption lines as proxies for the interstellar component in the absorption of \ion{C}{4}.  We do this two conservative ways. We first subtract an estimate of the contribution to the equivalent width from interstellar absorption using the largest equivalent width for the high-ionization line \ion{Si}{4}, which comes from the \ion{Si}{4}$\lambda$1393 line in both samples. We then calculate the equivalent width ratio of candidates to non-candidates, using these ISM-corrected stellar \ion{C}{4} equivalent widths. This result, with the propagated errors on the equivalent width measurement, is shown as purple horizontal lines in the left panel of \ref{metrat}. We repeat the ISM-corrected equivalent width ratio using a more conservative estimate of the ISM contribution, by using the largest equivalent width of all ISM lines we measure, that of the blend \ion{O}{1}/\ion{Si}{2}$\lambda$1303 (orange). We find that in both cases, the ratio of equivalent widths imply that candidates are more metal enriched relative to non-candidates, by at least a factor of 2, and quite possibly larger, in agreement with the findings from the emission component.

  \section{Discussion}
\label{Discussion}

The observations that we have presented are concordant in showing that 
 1) the candidates exhibit more energetic feedback than the non--candidates in the form of more turbulent bulk motions in the ISM; 2) the candidates have more
intense \Lya\ emission 
along with strong interstellar absorption, significantly deviating from the general trend of LBGs, pointing
to a distribution of \ion{H}{1}\ that is more conducive to the escape of
\Lya\ photons than in the non candidates; 3) the candidates have larger
metallicity than the non--candidates. We now discuss these three lines of
evidence in more detail.

\subsection{Evidence for differences in feedback in the ISM}

Figures \ref{lya} and \ref{ewew} show that on average the candidates exhibit strong,
redshifted Ly$\alpha$ emission, in combination with large equivalent width of
interstellar absorption lines, characteristics which strongly deviate from the
overall trend observed in LBGs at $z\sim 3$. Strong \Lya\ emission may escape
a star--forming galaxy if the ISM has patchy distribution around the
production regions, i.e. if there are holes punched in the neutral medium by
ionizing sources,
 or, if the line is generated in the receding part of the
outflow and is redshifted to an extent that its wavelength is too long for
resonant scattering by the intervening ISM. In this section, we discuss the evidence for each of
 these processes in our samples, and the implications they have for energizing the ISM.

\subsubsection{Holes in neutral gas and covering fraction}
Evidence of holes in the ISM of very compact star--forming
 galaxies has been reported by \citet{Heckman2011},
who have studied local analogs to LBGs with high-resolution UV spectroscopy.
The authors found that among galaxies that exhibit significant leakage of \Lya, and also of
ionizing radiation, 75\% also contain a very compact star-forming region (identified as a dominant compact object; DCO).
The interpretation is that holes in the ISM, presumably created by
feedback in the dense environment, allow both ionizing and \Lya\ photons to
escape. As a consequence, they also observed that the stellar continuum partially filled in absorption lines. In
these cases, the \Lya\ emission is observed as either a single peak \Lya\ line with significant flux 
blueshifted relative to the resonant wavelength of the line, or, as a
double--peaked line with the secondary peak located blue ward of the resonance
wavelength. Between 30 and 75\% of the total emission equivalent width of the
line is contributed by wavelengths bluer than resonance. Galaxies with no
filled--in interstellar lines and no leaking ionizing radiation have no
\Lya\ emission at wavelengths bluer than the resonance one.

Our candidate sample is similar to that of the DCO, in that their  $\Sigma_{SFR}$ are of similar order (the DCO have SFR $\sim$ 10$^{1}$ M$_{\odot}$yr$^{-1}$ and radii $\sim$100pc), and velocities from ISM line centroids in excess of 700 \kms.
In both stacks, despite being redshifted relative to the resonance wavelength by several hundred
\kms\, we
still observe some \Lya\ emission at wavelength bluer than the resonance one.
However, we found a lower fraction of the emission equivalent width is
contributed by these blue wavelengths compared to the galaxies in
\citet{Heckman2011}. Specifically, we find 7\% for the candidates and 15\% in the
non-candidates, which suggests that the distribution of \ion{H}{1}\ around the
regions of \Lya\ production has holes.

We do not observe any obvious
filling--in of the interstellar lines as observed in the spectra of the leaking DCOs, although this could be an effect of the
lower resolution of our spectra compared to those by \citet{Heckman2011}. In
low resolution spectra such as ours, however, the equivalent width of
saturated absorption lines, like the low--ionization metal lines that we
are considering here, does not depend on the column density of the
absorbing trough but on  the velocity spread of the individual clouds and the covering fraction. Thus,
it is possible that any fill--in of the lines is diluted at our resolution and
lost in the general velocity field of the trough.   Although the
  absorption lines are stronger in the candidates,  the physical interpretation of this fact isn't necessarily clear given the
  degeneracy between velocity spread and covering fraction.

Whether or not this difference in \Lya\ properties between candidates and non-candidates is due to a difference in the neutral gas covering fraction is difficult to say. In both samples, the fraction of blue shifted \Lya\ is uncertain due to the uncertainties in the systemic redshift estimates. In the case of the candidates, the majority of the \Lya\ emission is strongly redshifted, which may indicate that the emergence of the \Lya\ flux is modulated significantly by outflows. We will discuss evidence for this point in depth in Section 4.1.2. 
Nevertheless, a difference in the \ion{H}{1}\ covering fraction between our samples is
qualitatively consistent with the morphologies shown in Figure \ref{cutouts}. This figure illustrates the differences in morphology between our samples with rest-frame UV (tracing star-forming regions) and optical color (tracing the stellar distribution) composites. Non-candidates generally appear
to be less concentrated, with more extended and disturbed morphologies, and
most importantly often show the presence of UV flux from unobscured
star-forming knots. Meanwhile, the candidate sample are
characterized by smaller half-light radii and steeper Sersic indices (n$>$2)
\citep{Williams2014}, indicative of dense bulge-like stellar distributions. Thus, it is worth noting that the differences we observe in the ISM as probed by the \Lya\ properties could be related to morphology.
 Previous studies have found links between ISM
properties, outflows and morphology. 
 \citet{Law2012b} found that larger LBGs in general may exhibit a
decreased gas evacuation efficiency from feedback based on their measurements
of less strongly ionized outflows, and larger optical depth of gas at the
systemic redshift of the galaxies. They also found the largest absorption line velocity offsets from
the smallest galaxies. Our results are
qualitatively in agreement with this finding.

\subsubsection{Redshifted \Lya\ }

The stronger interstellar absorption of the candidates implies a larger spread
of velocities in the absorbing trough, which in turn suggest that their
gas kinematics is more turbulent than the non--candidates, likely as the result of a
more extreme feedback in their denser environment. This is also in qualitative
agreement with the faster velocity spread of the outflowing component of the candidates. The
relatively exceptional ISM of the candidates is illustrated in Figure
\ref{ewew}, which shows that while the equivalent widths of \Lya\ emission
and interstellar line absorption of the non--candidates follow the general
trend observed for LBG at $z\sim 3$-4, the candidates markedly deviate from it
because their \Lya\ emission is too strong given the amount of interstellar
absorption. In fact, given the equivalent width of the interstellar lines of
the candidates, their \Lya\ should have been observed in absorption had they
followed the general trend.

\begin{figure*} [!t]
\begin{center}
\includegraphics[scale=.4]{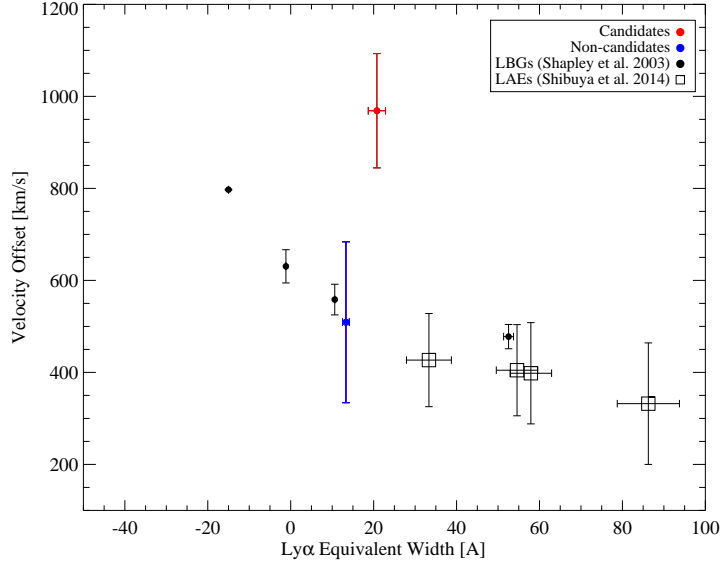}
\caption{The relationship between equivalent width of \Lya\ and velocity offset between \Lya\ and interstellar absorption lines. As equivalent width of \Lya\ decreases the velocity offset increases. The candidates lie significantly off this trend. In black are data reproduced from \citet[][LBG at z$\sim$3; circles]{Shapley2003} and \citet[][\Lya\ emitters; LAEs; squares]{Shibuya2014}.}
\label{lya_veloff}
\end{center}
\end{figure*}

Another evidence that the \Lya\ emission from the candidates is, on average,
significantly different from that of the general LBG population is illustrated
in Figure \ref{lya_veloff} \citep[adapted from ][]{Shapley2003, Shibuya2014}. This figure shows another trend characteristic of both LBGs and also Lyman-alpha emitters (LAEs) at 2$<$z$<$3, which is that the velocity offset between \Lya\ and ISM absorption lines decreases as the EW of the
\Lya\ increases. Galaxies with stronger \Lya\ emission
exhibit lower velocity offsets, whereas galaxies with \Lya\ in absorption
exhibit the larger velocity offsets. Other studies find similar trends
\citep{Vanzella2009, Jones2012, Song2014}, 
although \citet{Berry2012} do not find this
correlation very significant on an individual galaxy basis, presumably due to the larger scatter in the measurements from individual spectra. Our two average spectra show that
while the non--candidates are in very good agreement with this general trend,
the candidates strongly deviate from it, in the sense that their average
velocity offset (the mean value of all lines from Figure \ref{veloff} is
$\sim$970 \kms) is much larger than that of the other galaxies given the
observed \Lya\ strength. Again, given the observed velocity spread, the
\Lya\ of the candidates should have been observed in absorption had the
general trend be followed\footnote[1]{We noted in Section \ref{syst} that were
  we to adopt other methods of deriving systemic redshifts, we would still
  arrive at all our main conclusions.
  This is not entirely true in the case of the \Lya\ velocity offset relative to the interstellar absorption lines;
  using the method of \citet{Adelberger2005}, the mean velocity
  offset 
   measured from the candidate
  stack decrease to $\sim$560 \kms, and is identical to non-candidates. This
  is expected, as this method assumes all our galaxies have the same average
  velocity offsets of their galaxy sample, and ignores inherent differences
  between our two samples. We still detect faster velocity spread in the outflowing component from the ISM lines
  themselves compared to non-candidates using the Adelberger method, however,
  the candidates' deviation from the overall LBG trend in \Lya\ equivalent
  width and velocity difference is smaller. However, the high equivalent width
  of \Lya\ remains.}.
We suggest that winds accelerated to very high speed very close to the galactic center, in combination with fast outflows could produce the deviation from the global trend of LBGs.

The redshifted, high equivalent width
\Lya\ seen among candidates could also be a result of \Lya\ scattering off HI in
the immediate circumgalactic medium, and in particular, the outflowing
gas. Diffuse \Lya\ halos have been seen among LBG at z$\sim$3, regardless of
their spectroscopically determined \Lya\ equivalent width \citep[i.e. emission
  or absorption;][]{Steidel2011}. These diffuse halos are faint relative to
the UV continuum emitted by the central galaxy but appear to be generically
present, independent of the morphological properties of LBG. In some cases,
the total \Lya\ flux from halo regions between 2-4 arc seconds around the
galaxies is similar to that being emitted from the central 2" \citep[see ][]{Hayashino2004}.  The
main driver of the \Lya\ emission in the halos is likely the fraction of
photons that are able to escape the central regions \citep{Steidel2011}.

The faintness of these diffuse halos can to some extent be attributed to a
decreased density of the outflowing neutral gas with radius as the outflow
increases volume, and the fact that escaping \Lya\ photons end up distributed
over the much larger projected area of the halo. 
A simple model of such
outflowing media was outlined in \citet{Steidel2010}, who showed that
scattering of \Lya\ photons off circumgalactic gas in a high velocity,
spherically symmetric outflow with a uniform velocity field (i.e. independent
of galactocentric radius) and a covering fraction which is radially dependent
could explain the redshifted \Lya\ emission seen in LBGs. In this
context, it is possible that the high-velocity outflowing medium immediately
surrounding the candidate galaxies at small galactocentric radius could be
capable of redshifting the \Lya\ we observe.
 As discussed in
\citet{Steidel2011}, if strong \Lya\ is observed in combination with strong
interstellar absorption, the two must be originating from physically distinct
regions, or the \Lya\ would be modified \citep[e.g. the global trend in Figure \ref{ewew} seen
  by][]{Shapley2003, Jones2012}. It is possible that, due to the strong outflowing medium in
combination with the compact sizes of our galaxies, that dense, outflowing circumgalactic
material at small galactocentric radius may scatter sufficient \Lya\ to
reproduce the \Lya\ profiles we observe.

Mapping scattered light from outflowing material outside the ISM of the galaxy
may confirm or reject this interpretation either through narrow band imaging
or spectroscopy \citep[e.g.][]{Steidel2011, Martin2013}. We conclude based on
this data that the deviations of candidates from the trends of LBGs in terms of
their \Lya\ properties and other ISM properties is an indication of  a larger fraction of gas moving at higher speed, suggestive that feedback is more effective in compact LBGs.

 \subsubsection{Implications}

Taken all together, the larger bulk velocity of ISM gas in the candidates, their
larger velocity field and/or covering fraction as traced by the strength of the interstellar
absorption, the presence of significant \Lya\ emission blue ward of the line
resonant wavelength as well as the overall strength of the \Lya\ emission
lines, delineate a scenario where the ISM of these galaxies is subject to a
more extreme form of feedback compared to the non--candidates and the general
LBG population. This feedback would accelerate galactic winds to higher speed
at shorter distance from the galaxy center than in non--candidates (so that the
surface brightness of the \Lya\ line remains high and contributes to observed
flux) and create a patchy distribution of neutral hydrogen such that
\Lya\ photons can escape, including those close to the resonance wavelength or
even blue ward of it (the presence of which would need to be confirmed with higher dispersion spectra). The more extreme feedback also produces a generally
more turbulent ISM. Since the candidates differ from the non--candidates for
being more compact, and having higher surface density of star-formation, their more extreme
feedback might be the result of the denser environments where star formation
takes place. We will discuss this point in depth in Section 4.3.

\subsection{The origin of faster winds among candidates}

In Section \ref{Results} we presented evidence that the compact LBG studied here have
fast  
(weighted mean velocity -426 km s$^{-1}$) blueshifted velocity offsets measured from ISM absorption lines, with
 significantly larger equivalent width
seen among candidates.  Understanding the origin of the large velocity spread in the wind is
important in the context of star-formation quenching. The presence of these
strong winds in candidates, which are clearly more powerful than those in more
normal SFGs of similar mass at similar epoch, i.e. the non-candidates (weighted mean velocity -262 km s$^{-1}$), is an
indication that the conditions in the ISM of candidates may be more hostile to
cold gas, and lends support to the hypothesis of \citet{Williams2014} that the
candidates may quench sooner than normal SFGs. In this section, we discuss the plausible origins for this outflowing component.

 Conventionally the generation of fast winds ($\sim$1000 \kms) have been attributed to AGN \citep{Trump2006,Tremonti2007,Sturm2011,RupkeVeilleux2013,GaborBournaud2014, Genzel2014}, but see also
\citet{Coil2011}.
For candidates, the properties of emission lines from our limited data do not
indicate evidence of a detectable AGN. The rest-frame UV spectroscopic
properties of AGN among LBGs at z$\sim$2-3 have been studied by
\citet{Steidel2002, Hainline2011}, and do not share similar properties with
our candidate sample. If an AGN is present but obscured, this could prevent
its presence from being detected in the \ion{He}{2} / \ion{C}{4} emission line ratio. In a
recent analysis of a compact post-starburst galaxy from the sample of
candidate progenitors of local passive galaxies \citep{Marchesini2014},
\citet{CemileMarsan2014} do find evidence of an obscured AGN using rest frame
optical emission lines. Their galaxy, like ours, is not X-ray detected, and
has a similarly small \ion{C}{4}\ / \ion{He}{2} emission line ratio ($\sim$0.8).  However,
\citet{Rupke2005, Krug2010} found that obscured AGN do not launch high-speed
large-scale outflows, and rather only launch local outflows near the central
black hole. Further analysis including restframe-optical spectroscopy will be
necessary to investigate further the presence of AGN and its effect in driving
winds in compact LBGs.

Whether or not star-formation driven (or supernova driven) winds are capable
of halting star-formation on galaxy scales is still a matter of debate. Starburst driven winds often occur with lower velocities \citep[$\sim$500\kms, e.g.][]{Fabian2012}.  However, evidence that star-formation can launch high velocity winds is accumulating, especially at high SFR surface densities, 
\citep{Heckman2011, DiamondStanic2012, Bradshaw2013, Sell2014}.

Our candidates have on average larger SFRs (although overlapping distributions in SFR), but have distinctly higher surface density of star-formation (see Figure \ref{sampleprop}). 
As discussed in Section 4.1, their surface density of SFR are similar to those of the DCO, whose outflows were determined to be star-formation driven \citep{Overzier2009, Heckman2011}. 
Therefore it is possible that the surface density of star-formation is related to our detection of differences in outflow velocities. (Nevertheless, as we showed in Section 3, our data is also consistent with the scaling relation presented in \citet{Martin2005} between velocity and SFR.) 
Based on conflicting evidence in the literature, it is unclear if outflow velocities in SFGs are universally correlated with SFR or SFR surface density \citep{Martin2005, Rupke2005, Erb2012, Weiner2009, Sato2009, Rubin2010, Steidel2010, Chen2010, Kornei2012, Law2012b, Bordoloi2013, KSLee2013, Song2014}. With this data it is not possible to identify whether star-formation or surface density of star-formation is more strongly correlated with gas velocity in our galaxies.

 Our detection of significant populations of massive evolving stars (i.e. the WR population), which currently affect the ISM with their strong winds, is consistent with this idea that the outflowing gas we detect are star-formation driven. The feedback from radiation and winds of the massive stars themselves (while on the main sequence) is especially significant at high stellar surface densities, leading \citep{Hopkins2010} to conclude that feedback from massive stars sets a limit on stellar surface densities in galaxies \citep[$\Sigma_{max}\sim$10$^{11}$M$_{\odot}$kpc$^{-2}$; similar to that observed in z$\sim$2 ultra-compact passive galaxies][]{Cassata2011,Cassata2013}. These WR stars are expected to form SNe on short timescales.  
It has been shown that SNe from concentrated groups of young stars, such as that in large star-clusters, may be more effective at evacuating gas than isolated SNe \citep[e.g.][]{Murray2011, NathShchekinov2013, Sharma2014}. Therefore, in such compact galaxies as our candidate sample, the compact stellar configurations of both the massive stars and SNe could thus be very effective at evacuating gas, compared to more extended distributions of stars.

  \subsection{Evidence for differences in metallicity}
\label{relmet}

The implication that candidates exhibit higher metallicities may be important for the efficacy of feedback in the near future (i.e. over the next Gyr) as these galaxies evolve. 
If a significant portion of the feedback in these galaxies comes from a dense concentration of high-mass stars with strong stellar winds affecting a very concentrated distribution of cold ISM gas, this feedback may increase its impact as the metallicity of the galaxy continues to increase. 

The strength of winds from WR stars in particular, the features of which we have observed in the candidate sample, is dependent strongly on metallicity \citep{VinkdeKoter2005}, with wind velocity and mass loss (and therefore momentum) increasing with increasing metallicity \citep[e.g.][]{Leitherer1992, KudritzkiPuls2000,Crowther2000,NugisLamers2000, CrowtherHadfield2006,Crowther2007, Mokiem2007}.
Over time, therefore, as more metal enriched young stars produce stellar winds, the energy deposition to the surrounding dense ISM is expected to increase.

Although calibrating our metallicity measures onto an absolute scale is not
possible with our data, and therefore it is unclear that candidates are
already at metallicities high enough to be consistent with those of passive galaxies, the
suggestion of enhanced metallicities does provide evidence for a more advanced
evolutionary state. It is not currently possible to identify differences in the star-formation timescale of the two samples. Future robust measurements of metallicities and abundance ratios may be able to  differentiate between a
scenario where the candidate sample form stars on a faster timescale and
quench earlier, or, form earlier on a similar timescale (but formed stars
longer).

The presence of
strong WR features, which arise during relatively short phases (10$^{5}$ yrs)
relative to the much larger timescale probed by the stack, argues for a
relatively continuous star-formation history, perhaps driven by cold
accretion, rather than burst-like star-formation from mergers which could elevate their star-formation
during short periods of time. We argue that that the enhanced metallicity of
the candidates relative to the non--candidates is consistent with the idea 
that the former form their stars on a faster timescale than the latter 
during a period of sustained dissipative accretion of gas.
Direct constraints on this interpretation will be possible with future NIR
spectroscopic observations that would not only allow robust systemic redshifts
to be measured, but also metallicity measurements, including $\alpha$/Fe
abundances, which could directly test our interpretation that these galaxies
follow a faster evolution than normal main sequence galaxies.

\subsection{Are The Candidates About To Quench Star Formation?}

As detailed in \citet{Williams2014}, the candidates have been selected based
on their morphology, stellar mass and star--formation rate to reproduce the
observed properties of the CPG at $z\sim 2$, assuming that {\it they quench
  their star formation activity} shortly after the epoch of observation. Is
this assumption well posed and is there any evidence that supports it?

We have provided evidence that the kinematics of the ISM of the
candidates shows that they are subject to a more energetic feedback than the
non candidates. The half--light radius of the candidates is $\approx 2\times$
smaller than the non--candidates, and since their average stellar mass is the
about the same, the stellar density is $\approx 8\times$ higher. The velocity
 centroid of the outflowing component of the gas in candidates also is $\approx 2\times$ faster. Thus, the
mechanical energy density of the ISM of the candidates is roughly $32\times$
larger than that of the non candidates. Since the star--formation rate of the
candidate also is larger by about $2\times $, the radiation energy density
contributed by massive stars is also larger by the same amount.

The larger energy deposited in the ISM of the candidates is likely to increase
its turbulence, temperature and ionization state, and this is broadly
consistent with the observed \Lya\ properties, strength of the IS absorption
lines and presence of \ion{He}{2}\ emission. The larger star--formation rate
is also likely to result in large metallicity, which is also consistent with
the stronger \ion{C}{4}\ emission of the candidates. Their larger
star--formation rates would also produce more dust, except that the larger
radiation field and higher ISM temperature are likely to be conducive to
destruction of dust grains more effectively than in the non--candidates, a
scenario that is supported by the far--IR observations which show that
dust--reprocessed emission is detected among the non--candidates but not among
candidates, despite their larger SFR \citep{Williams2014}.

There is no evidence for AGN activity either in the X--ray emission or in the
UV spectra of the candidates (as well as the non-candidates). So, if the quenching has
to come by feedback, then this is likely to be provided by star formation
through heating and removal of gas. The outflow velocity of both candidates
and non candidates are very likely larger than the escape velocity from the
regions where the stars are forming, $v_{esc}\sim \sqrt{2\, G\,
  M_{star}/r_e}\approx 300$ \kms\ for the candidates and $\approx 200$
\kms\ for the non candidates if the dark matter gravity can be neglected
within the radius where star formation is observed. We have no direct
information about the mass loading of the outflows, but if it is roughly
proportional to the star formation rate \citep{Heckman2011}, then over a
similar period of time, e.g. from $z\sim 3$ to $z\sim 2$, the candidates will
have ejected one order of magnitude more gas in $\approx 1/5$ of the time than
the non candidates, because their SFR is higher ($2.5\times$), the velocity of outflowing component of ISM gas is larger ($2\times$ )and their size is smaller ($2\times$). Combined
with heating the gas to a larger temperature, this might be sufficient to
quench them. 

Thus, it seems to us physically motivated to think that the candidates might quench
their star formation rate sooner and in a shorter time scale than the non--candidates.  This supports the validity of the selection criteria (which was based solely on morphology, stellar mass and starÐformation rate) that we adopted to identify them as progenitors of z $\sim$ 2 CPGs.
 Unfortunately, we are not aware of any way at this time to
quantitatively predict, based on the observed properties of the candidates,
e.g. their $\approx 30\times$ energy density relative to the
non--candidates, when and how fast they will quench.  An additional question is whether the feedback is sufficient to prevent ejected gas recycling, such that cooling gas is prevented from refueling future star-formation.
 Future observations at
(sub)millimeter wavelengths of the dust content and cold gas morphology and
kinematics, e.g. with ALMA, as well as future refinement of theoretical
modeling of feedback, will provide a path to progress.

\section{Summary}
We have presented rest-frame UV spectroscopy of a sample of LBGs which may be the progenitors of the high-redshift CPGs, compared with that of more normal LBGs (in terms of the mass-size relation). We find evidence for faster outflowing gas among these candidate progenitors relative to non-candidate LBGs as traced by blue shifted interstellar absorption lines. These lines also have higher equivalent widths, indicative of a higher velocity dispersion among intervening clouds in the ISM. This strong interstellar absorption is seen in combination with strong, redshifted \Lya\ which may be explained by strong outflows. Analysis of emission lines present in the rest-frame UV spectrum do not suggest the presence of an AGN, and we conclude that the outflows are star-formation driven. We see significant features due to massive stars in candidates, in particular those of metal rich WR stars (\ion{C}{4}\ and \ion{He}{2}\ ), and argue that this is evidence that candidates are more metal enriched than normal, extended LBGs at that epoch. This could be a result of an earlier formation time, or a more rapid evolutionary timescale. 
 We conclude that these plausible progenitors are distinct from normal LBGs (at the same mass and epoch) and further investigation of this sample, in particular with larger samples and NIR spectroscopy, will put important constraints on quenching mechanisms affecting compact galaxies at high-redshift.
 
 \acknowledgments
 We thank the anonymous referee whose valuable suggestions have improved this paper significantly. We thank Naveen Reddy and Charles Steidel for graciously providing the HIRES spectrum of D6. We also thank Alvio Renzini for reading this document and providing his valuable and insightful comments.
 CCW thanks Shawn Roberts, Joseph Burchett, Todd Tripp, Joseph Meiring, Claudia Scarlata, Eric Gawiser, Nadia Zakamska, Sanchayeeta Borthakur, Brandon Bozek for enlightening discussions.  
 This work is based on observations taken
by the CANDELS Multi-Cycle Treasury Program with the NASA/ESA HST.  We
acknowledge support from grant program NSF AST 08-8133, and support for {\it
  HST} Program GO 12060.10-A was provided by NASA through grants from the
Space Telescope Science Institute, which is operated by the Association of
Universities for Research in Astronomy, Inc., under NASA contract NAS5-26555.

\bibliographystyle{aa}

\end{document}